\documentstyle[psfig,paspconf]{article}
\def\mass#1{${\mathrm{#1\,M}_\odot}$}
\newcommand{\chem}[2]{$\rm{}^{#1}\kern-0.8pt#2$}
\newcommand{\chim}[2]{\rm{}^{#1}\kern-0.8pt#2}
\newcommand{\reac}[6]{$\rm\,{}^{#1}\kern-0.8pt{#2}\,({#3}\,,{#4})\,
           {}^{#5}\kern-0.8pt{#6}\,$}
\newcommand{\gsimeq}{\,\,\raise0.14em\hbox{$>$}\kern-0.76em\lower0.28em\hbox  
{$\sim$}\,\,}
\newcommand{\lsimeq}{\,\,\raise0.14em\hbox{$<$}\kern-0.76em\lower0.28em\hbox  
{$\sim$}\,\,}

\newcommand{\ms}{$\rm M_{\odot}$}

\begin{document}

\title{Some selected comments on cosmic radioactivities}

\author{Marcel Arnould$^1$, Georges Meynet$^2$ and Nami Mowlavi$^2$}

\affil{$^1$ Institut d'Astronomie et d'Astrophysique, Universit\'e Libre de
            Bruxelles \\
            C.P. 226, Bd. du Triomphe, B-1050 Brussels, Belgium\\
       $^2$ Observatoire de Gen\`eve, CH-1290 Sauverny, Switzerland}
 
\begin{abstract}
Radionuclides with half-lives ranging from some years to billions of years
presumably synthesized outside of the solar system are now recorded in `live'
or `fossil' form in various types of materials, like meteorites or the galactic
cosmic
rays. They bring specific astrophysical messages the deciphering of which is
briefly
reviewed here, with special emphasis on the contribution of Jerry Wasserburg

First, the virtues of the
long-lived (half-lives $t_{1/2}$ close to, or in excess of
$10^9$ y) radionuclides as galactic chronometers are discussed in the light of recent
observational and theoretical works. It is concluded that the trans-actinide clocks
based on the solar system abundances of \chem{232}{Th}, \chem{235}{U} and
\chem{238}{U} or on the \chem{232}{Th} surface content of some old stars are still
unable to meaningfully complement galactic age estimates derived from
other independent astrophysical methods. In this respect, there is reasonable hope
that the \chem{187}{Re}-\chem{187}{Os} chronometric pair could offer better
prospects. The special case of \chem{176}{Lu}, which is a pure s-process product, is
also reviewed. It is generally considered today that this radionuclide cannot be
viewed as a reliable s-process chronometer.

Second, we comment on the astrophysical messages that could be brought by short-lived
($10^5 \lsimeq t_{1/2} \lsimeq 10^8$ y) radionuclides that have been
present in live or in fossil form in the early solar system. From an astrophysical
point of view, the demonstrated early existence of live short-lived radionuclides is
generally considered to provide the most sensitive radiometric probe concerning
discrete nucleosynthetic events that presumably contaminated the solar system at
times between about $10^5$ and $10^8$ y prior to the isolation of the solar material
from the general galactic material. Of course, this assumes implicitly that the
radionuclides of interest have not been synthesized in the solar system itself. This
is still a matter of debate, as we briefly stress. If indeed the short-lived
radionuclides that have been present live in the early solar system are not of local
origin, the external contaminating agents that have been envisioned are supernovae,
evolved stars of the Asymptotic Giant Branch (AGB) type, or massive mass-losing
stars of the Wolf-Rayet (WR) type. We comment on some aspects of the AGB or WR
contamination. In the latter case, we discuss more specifically the role of rotation
and of binarity on the predicted yields of
\chem{26}{Al}, a radionuclide of special cosmochemistry and astrophysics interest.
Some comments are also devoted to
\chem{146}{Sm} and \chem{205}{Pb}. The former one is a short-lived p-process
radionuclide that has most probably been in live form in the solar system, while the
latter one is of s-process origin. It is shown to raise interesting nuclear physics
and astrophysics questions, and to deserve further cosmochemical studies in order to
evaluate its probability of existence in live form in the early solar system.

Third, the case of extinct short-lived radioactivities
carried by pre-solar grains is shortly mentioned, and some comments are
made about the possible origin of these grains. 

Finally, a brief mention is made to $\gamma$-ray line astrophysics, which provides
interesting information on live short-lived radionuclides in the present interstellar
medium, and thus complements in a very important way the study of extinct
radionuclides in meteorites. This is illustrated with the case of \chem{26}{Al}.
\end{abstract}

\vfill\eject

\section{Introduction}

Since its discovery about one century ago, nuclear radioactivity
has been the focal point of a vast amount of fundamental and applied research.
With
time, it has become clearer and clearer that some radionuclides even try to
tell us
something quite exciting about certain facets of the Universe in general, and
about
our solar system in particular. With his legendary enthusiasm and expertise,
Jerry
Wasserburg has searched without respite for the Rosetta Stone leading to the
understanding of the highly complex hieroglyphic message delivered by
radionuclides
to astrophysicists and cosmochemists. This contribution is a
tribute to this aspect of Jerry's mutiple activities. We will in particular
briefly
review some selected aspects of nucleo-cosmochronology, as well as of the field
of
the isotopic anomalies of radionuclidic origin. Additionally, the decay of
certain
radionuclides manifests itself quite spectacularly through the emission of
de-excitation lines in the
$\gamma$-ray domain. The study of these radionuclides offers invaluable
information
on their production sites and is the subject of
$\gamma$-ray line astronomy, a recently developed astrophysical discipline
which has
not escaped Jerry's intellectual curiosity. 

%In contrast, we will not deal here with the very interesting subject of the
%production of a large variety of radionuclides in terrestrial or solar-system
%extra-terrestrial matter bombarded by galactic cosmic rays or solar
%energetic particles. This spallative production is of importance for the deciphering
%of the record of these energetic particles, and in particular of the time variations
%of their fluxes, but its interest goes well beyond astrophysics or
%planetology. The reader is referred to e.g. Michel (1998) for a review and
% some references.

\section{Cosmochronometry}
 
The dating of the Universe and of its various constituents, referred to as
`cosmochron\-ology', is one of the tantalizing tasks in modern science. 
This field is in fact concerned with different ages, each  one of them
corresponding to
an epoch-making event in the past (e.g.~Vangioni-Flam et al. 1990 for many 
contributions on this
subject). They are in
particular the age of the Universe $T_{\rm U}$, of the  globular clusters
$T_{\rm
GC}$, of the Galaxy [as (a typical?) one of many  galaxies] $T_{\rm G}$, of the
galactic disc
$T_{\rm disc}$, and of the  non-primordial nuclides in the disc $T_{\rm nuc}$,
with
$T_{\rm U}
\gsimeq  T_{\rm GC} \approx$ ($\gsimeq$?) $T_{\rm G} \gsimeq T_{\rm disc}
\approx
T_{\rm nuc}$. As a consequence, cosmochronology involves not only
cosmological models and observations, but also various other astronomical and
astrophysical studies, and even invokes some nuclear physics information. 

The cosmological models can help in determining  $T_{\rm U}$, as well as, to some
extent at least, 
$T_{\rm GC}$ and $T_{\rm disc}$ (e.g. Fowler \& Meisl 1986, Tayler 1986, Clayton
1988, Arnould \& Takahashi 1990a). The $T_{\rm GC}$ or $T_{\rm disc}$ values have
also been evaluated from
the use of the Herztsprung-Russell diagram (HRD)
(e.g.~Jimenez 1998, VandenBerg et al. 1997
for recent reviews), or of so-called `luminosity functions' which provide the
total number of stars per absolute magnitude interval as a function of absolute
magnitudes. In particular, the luminosity function of white-dwarf stars has
been
proposed as a priviledged $T_{\rm disc}$ evaluator 
(e.g. Winget et al. 1987, Hernanz et al. 1994, Oswalt et al. 1996).
 
Nucleo-cosmochronological techniques have also been
developed in order to evaluate
$T_{\rm nuc}$, and are briefly discussed below. 
Each of these methods has advantages and weaknesses of its own, as briefly
reviewed by e.g.~Arnould \& Takahashi (1990a). The age estimates they provide are
sketched in a synopsis form in Fig.~1, which also displays the limits derived from
the nucleo-cosmochronological approach briefly discussed below. 
 
\vskip1truecm
\centerline{EDITOR: FIGURE 1 HERE}
\vskip1truecm

\subsection{`Long-lived' nucleo-cosmochronometers: generalities} 

The dating method that most directly relates to nuclear astrophysics is 
referred to as `nucleo-cosmochron\-ology.' It primarily aims at determining
the age $T_{\rm nuc}$ of the nuclides in the galactic disc through the use of
the observed bulk (meteoritic)  abundances of radionuclides with lifetimes
commensurable with presumed $T_{\rm disc}$ values (referred to in the following
as
`long-lived' radionuclides). Consequently, it is hoped to provide at least a
lower
limit to $T_{\rm disc}$. The most studied chronometries involve \chem{187}{Re}
or the
trans-actinides \chem{232}{Th}, \chem{235}{U} and \chem{238}{U}.    

In order to establish a good chronometry based on these radioactive nuclides,
one
needs to have firstly a good set of input data concerning (isotopic) abundances
and nucleosynthesis yields, in addition to the radioactive half-lives. Another
issue
concerns the necessity, and then the possibility, of using detailed models for
the
chemical evolution of the Galaxy in order to gain a reliable
nucleo-cosmochronological information if indeed the bulk solar-system
composition
witnesses the perfect mixing of a large number of nucleosynthetic events. The
status
of these various requirements is briefly examined in the following sections for
several cosmic clocks.
\subsubsection{The trans-actinide clocks} 

The familiar long-lived   
\chem{232}{Th}-\chem{238}{U} and \chem{235}{U}-\chem{238}{U} chronometric pairs
(Fowler \& Hoyle 1960) are developed on grounds of their abundances at the time of
solidification in the solar system some $4.56 \times 10^9$ y ago. This
information
is obtained by extrapolating back in time the present meteoritic content of
these
nuclides. If the so-derived abundances are affected by some uncertainties,
these
are not, however, the main problems raised when attempting to use these
radionuclides as reliable nuclear clocks.  Their usefulness in this respect
indeed
depends in particular on the availability of precise production ratios. Such
predictions at the level of accuracy needed for getting a truly useful
chronometric
information are out of reach at the present time. One is indeed dealing with
nuclides
that can be produced by the r-process only, which suffers from very many
astrophysics and nuclear physics problems, in spite of much effort by many
researchers. The r-process problems are particularly acute
for the Th and
U isotopes referred to above. They are indeed the only naturally-occuring
nuclides beyond \chem{209}{Bi}, so that any extrapolation relying on
semi-empirical
analyses and fits of the solar r-process abundance curve is in danger of being
especially unreliable. 
This difficulty is illustrated by the recent calculations of Goriely \& Clerbaux
(1999). One has to note, however, that many authors express opposite views on this
question (e.g. Pfeiffer et al. 1998). An additional problem the trans-actinide
clocks has to face relates to the fact that
most of the r-process precursors of U and Th are unknown in
the laboratory, and will remain so for a long time to come.  Theoretical
predictions of properties of interest, like masses, 
$\beta$-decay strength functions and fission barriers, are extremely
difficult, particularly as relevant calibrating data are largely missing.
The problems mentioned above would linger even if a realistic r-process model were
given, which is not the case at the present time (e.g. Arnould \& Takahashi 1999).
Last but not least, most
of the tremendous amount of work devoted in  the past to the trans-actinide
chronometry   has adopted simple functionals for
the time dependence of the r-process nucleosynthesis rate with little
consideration
of the chemical evolution in the solar neighborhood. This view, which
originated almost 4 decades ago (Fowler \& Hoyle 1960), has had (and still has) a few
sympathisers indeed (e.g.
Cowan et al. 1991; also Arnould \& Takahashi 1999 for some references). 

The necessity of the development of the long-lived chronometers in the
framework of
models for the chemical evolution of the Galaxy has been first pointed out by
Tinsley (1977). The introduction of nucleo-cosmochronological
considerations in such models is not a trivial matter, however. The intricacies
come
in particular from `astration' effects, which have to do with the fate of the
chronometers once absorbed from the interstellar medium (ISM) by the stars at
their
birth (e.g. Yokoi et al. 1983). However, Jerry, in collaboration with Schramm
(Schramm \& Wasserburg 1970), has made an important contribution to nucleo-cosmochronology
by
showing that one can make the economy of these chemical evolution models as
long as a mere determination of {\it age limits} could satisfy one's curiosity.
This
interesting so-called `model-independent approach' has led to the conclusion
that $9 \lsimeq T_{\rm nuc} \lsimeq 27$ Gy (Meyer \& Schramm 1986).
 
There has also been an attempt to develop a Th-chronometry (e.g. Pagel 1997)
on grounds of the relative abundances of Th and Eu (which is
presumed to be dominantly produced by the r-process) observed at the surface of a
variety of stars with metallicities in a wide range of values (da Silva et al. 1990,
Fran\c{c}ois et al. 1993).\footnote{Originally, an attempt was made to use the
observed Th/Nd ratios (Butcher 1987), albeit the disadvantage of Nd being possibly
produced also by the s-process} Under the assumption, which may sound reasonable, but
has not at all to be taken for granted, that any r-process in the past has produced
Th and Eu with a constant solar-system ratio, the age determination is reduced to the
problem
of mapping the metallicity to time through a chemical evolution model
(e.g. Pagel 1997). This is by far not a trivial matter. One of the difficulties
noted by Fran\c{c}ois et al. (1993) arises from the complex trend of the
observed Th/Eu abundance data with metallicity. The Th observations in highly
metal-deficient stars (e.g. Sneden et al. 1998, and references therein) alleviate the
difficulties inherent in the chemical evolution models. The considered stars have
indeed been born in the Galaxy at times much shorter than the Th decay half-life. As a
consequence, Th could not have been affected by chemical evolution effects, and only
a simple exponential decay law has to be applied. This does not imply, however, that 
$T_{\rm nuc}$ can be trivially derived from these observations. One of the main
problems again concerns the assumption that the Th/Eu production ratio is strictly
solar in all r-process sites. There is at best some hint that this could indeed be
the case (Sneden et al. 1998), but, in our opinion, it would be premature to interpret
this hint as a demonstration.
  
The Th-chronometry could be put on safer grounds if the Th/U ratios  
would be known in a variety of stars with a high enough accuracy. These
nuclides
are indeed likely to be produced simultaneously, so that one may hope to be
able to
predict their production ratio more accurately than the Th/Eu one. Even in such
relatively favorable circumstances, one would still face the severe question of
whether Th and U were produced  in exactly the same ratio in presumably a few
r-process events (a single one?) that have contaminated the material from
which metal-poor stars formed. Even if this ratio turns out to be the same
indeed, its precise value remains to be calculated (see e.g.
Arnould \& Takahashi 1990b for an illustration of the dramatic impact of a variation
in the predicted Th/U ratio on derived ages).
\subsubsection{The \chem{187}{Re} - \chem{187}{Os} chronometry} 

First introduced by Clayton (1964), the chronometry using the
\chem{187}{Re}-\chem{187}{Os} pair is able to avoid the difficulties 
related to the r-process modeling. True, \chem{187}{Re} is an r-nuclide.
However, \chem{187}{Os} is not produced  directly by the r-process, but 
indirectly via the $\beta^-$-decay of \chem{187}{Re} ($t_{1/2} 
\approx 43$ Gy) over the galactic lifetime. This makes it in principle
possible to derive a lower bound for $T_{\rm nuc}$
from the mother-daughter abundance ratio, provided that the `cosmogenic'
\chem{187}{Os} component is deduced from the solar abundance by subtracting
its s-process contribution. This chronometry is thus in the first instance
reduced to a question concerning the s-process. Other good news come from the
recent progress made in the measurement of the abundances of the concerned
nuclides in meteorites (e.g.~Faestermann 1998 for references). This input
is indeed essential also for the establishment of a reliable chronometry.
 
Although the s-process is better understood than the r-process, this
chronometry is facing specific problems. They may be summarized as follows
(see e.g.~Takahashi 1998 for a short account): 1) the evaluation of the
\chem{187}{Os} s-process component from the ratio of its production to the one
of the s-only nuclide \chem{186}{Os} is not a trivial matter, even in the
simple local  steady-flow approximation (constancy of the product of the
abundances by the stellar neutron capture rates over a restricted
$A$-range). The difficulty relates to the fact that the \chem{187}{Os} 9.75
keV excited state can contribute significantly to the stellar neutron capture
rate because of its thermal population in s-process conditions ($T \gsimeq
10^8$ K) (e.g.~Winters et al. 1986, Woosley \& Fowler 1979). The ground-state capture rate
measured in the laboratory has thus to be modified by a theoretical
correction. In addition, the possible branchings of the s-process path in
the $184 \leq A \leq 188$ region may be responsible for a departure from the
steady-flow predictions for the \chem{187}{Os}/\chem{186}{Os} production ratio
(e.g.~Arnould et al. 1984, K\"appeler et al. 1991); and 2) at the high temperatures, and thus
high ionisation states, \chem{187}{Re} may experience in stellar interiors,
its $\beta$-decay rate may be considerably, and sometimes enormously, enhanced
over the laboratory value by the bound-state $\beta$-decay of its ground
state to the 9.75 keV excited state of \chem{187}{Os} (e.g.~Yokoi et al. 1983).
Such an enhancement has recently been beautifully confirmed by the
measurement of the decay of fully-ionised \chem{187}{Re} at the GSI storage
ring (Bosch et al. 1996, Kienle et al. 1998). The inverse transformation of \chem{187}{Os}
via free-electron captures is certainly responsible for additional
corrections to the stellar \chem{187}{Re}/\chem{187}{Os} abundance
ratio (e.g. Arnould 1972, Yokoi et al. 1983). Further complications arise
because these two nuclides can be concomitantly destroyed by
neutron captures in certain stellar locations (Yokoi et al. 1983).
 
All the above effects have been studied in the framework of detailed evolution
models for $1 \lsimeq M \lsimeq 50$ M$_\odot$ stars and of a galactic chemical
evolution model that is constrained by as many observational data in the solar
neighborhood as possible in order to reduce as much as possible the uncertainties that
are inherently associated with any model of this type (Takahashi 1998, Takahashi et
al. 1998). This work, which is an up-date of Yokoi et al. (1983) with regards to
meteoritic abundances, nuclear input data, stellar
evolution models and observational constraints, concludes that $T_{\rm nuc}
\approx 15
\pm 3$ Gy. Even lower ages of about 9 Gy, as derived from the model-independent
approach (Schramm 1990, Schramm \& Wasserburg 1970;  see Sect.~2.1), cannot
 conclusively be excluded
within the remaining uncertainties in the chemical evolution model parameters.

These results may imply that the \chem{187}{Re} - \chem{187}{Os} chronometry
has not
yet much helped narrowing the age range derived from other methods. There is
still
ample room for improvements, however, and there is reasonable hope that the Re
- Os
chronometry will be able to set some meaningful limits on $T_{\rm nuc}$ in a
near
future, and independently of other methods which suffer from problems of other types.
 
\subsubsection{\chem{\rm 176}{\rm Lu}, a long-lived s-process radionuclide}

The long-lived \chem{176}{Lu} ($t_{1/2} = 41$ Gy) has the remarkable property
of
being shielded from the r-process, and thus to be a pure s-process product.
Some
early works (Arnould 1973, Audouze et al. 1972) have proposed this radionuclide to be 
a potential chronometer for the s-process, the other long-lived radionuclides
probing the r-process instead. These studies pointed out some possible
uncertainties in the solar \chem{176}{Lu} abundance, as well as in its
production
predicted from s-process models. The latter problem relates directly to the
branching
in the s-process path due to the 125 keV \chem{176}{Lu^{\rm m}} isomeric state.
More
specifically, the two different paths
\chem{175}{Lu}(n,$\gamma$)\chem{176}{Lu^{\rm
g}}(n,$\gamma$)\chem{177}{Lu}($\beta^-$)\chem{177}{Hf} and
\chem{175}{Lu}(n,$\gamma$)\chem{176}{Lu^{\rm
m}}($\beta^-$)\chem{176}{Hf}(n,$\gamma$)\chem{177}{Hf} may well develop during
a
s-process (\chem{176}{Lu^{\rm g}} designates the \chem{176}{Lu} ground state).
The
resulting
\chem{176}{Lu^{\rm g}}/\chem{176}{Hf} production ratios depend on the relative
importance of these two branchings, and thus mainly on the relative population
of
the ground and isomeric \chem{176}{Lu} states. Two limiting situations are
relatively
simple to handle. The first one is obtained if \chem{176}{Lu^{\rm g}} and
\chem{176}{Lu^{\rm m}} have no time in a given astrophysical environment for
being
connected electromagnetically. This situation is made plausible by the 
large difference in the spin and $K$ quantum
number of the two states. In such conditions, the relative importance of the
two
s-process branches is just given by the \reac{175}{Lu}{n}{\gamma}{176}{Lu^{\rm
m}}/\reac{175}{Lu}{n}{\gamma}{176}{Lu^{\rm g}} cross section ratio, the value
of
which can be obtained from experiments. The other extreme is obtained if 
\chem{176}{Lu^{\rm g}} and \chem{176}{Lu^{\rm m}} are coupled
electromagnetically
strongly enough for the relative populations of these two states to be
`thermalized', i.e. follow the rules of statistical equilibrium. In such
conditions, the relative importance of the two s-process branches is
essentially
governed by temperature, as is the effective decay rate of the thermalized
\chem{176}{Lu}. 

Since the pioneering studies mentioned above, much work has been devoted to the
question of the possibility of thermalization of the \chem{176}{Lu} isomeric
and
ground states in astrophysical plasmas, and to the measurement of the neutron
capture cross sections needed for the calculation of the s-process
\chem{176}{Lu}/\chem{176}{Hf} production ratio (e.g. Klay et al. 1991, Lesko et al.
1991, Doll et al. 1999, and references therein). From these efforts, it is generally
concluded today that the \chem{176}{Lu^{\rm g}} s-process yields are so
sensitive
to temperatures and neutron densities that they cannot be evaluated precisely
enough
for chronological purposes. Instead, \chem{176}{Lu^{\rm g}} could rather be
considered as a s-process thermometer.
\section{The message from extinct `short-lived' radionuclides}

The discovery of isotopic anomalies attributed to the decay in some
meteoritic material of now extinct radionuclides with half-lives in the
approximate
$10^5 \lsimeq t_{1/2} \lsimeq 10^8$ y range (referred to in the following as
`short-lived' radionuclides) has broadened the original astrophysical interest
for
cosmic radioactivities. Even the `ultra-short' radionuclides \chem{22}{Na}
($t_{1/2}
= 2.6$ y) and \chem{44}{Ti} ($t_{1/2} \approx 60$ y; Wietfeldt et al. 1999)
are likely to have left
their signatures in some meteorites. The interpretation of the message from these
anomalies has been the focus of much work and excitement.

One important issue raised by the extinct radionuclides concerns their
presence in the early solar system in `live' form, or just in the form of their
daughter products (`fossils'). In the first case, the anomalies have of course
to be
located in solar-system indigenous solids, while they have to be found in alien
(presolar) material in the second situation. At present, there is clear
evidence that
meteorites contain both live and fossil signatures of short-lived nuclides, and
the messages they carry are quite different indeed. In contrast, the meteoritic
content of the ultra-short-lived nuclides has obviously to be of fossil nature,
in
view of the lifetimes involved. 

\subsection{Live short-lived radionuclides in the early solar system}
 
At the end of the sixties, Jerry and his collaborators (Schramm et al. 1970) have
contributed in an important way to the pioneering searches for the signatures
of
extinct radionuclides in meteorites (e.g. Wasserburg \& Papanastassiou
1982 for a historical
account) by establishing techniques for the high precision measurement of the
Mg
isotopic composition in order to search for \chem{26}{Mg} excesses due to the
\chem{26}{Al} decay in meteorites of different types and in lunar samples. From
this
study, it was concluded that the upper limits on the
(\chem{26}{Al}/\chem{27}{Al})$_0$ ratio\footnote{Here and in the following, the
subscript 0 refers to the start of solidification in the solar system some 4.56
Gy ago} in the analyzed materials was ranging from well below $10^{-6}$
to about $2 \times 10^{-6}$. It is established by now that \chem{26}{Al} has
been
live in the solar system at a canonical level of
(\chem{26}{Al}/\chem{27}{Al})$_0
\approx 5 \times 10^{-5}$
(MacPherson et al. 1995). Jerry and his collaborators have also contributed in a
significant way to the accumulation of persuasive experimental evidence for the
existence of other
live radionuclides. This concerns nowadays
\chem{53}{Mn},
\chem{60}{Fe},
\chem{107}{Pd}, \chem{129}{I}, \chem{146}{Sm} and \chem{244}{Pu}. Other likely
candidates are \chem{182}{Hf} and \chem{41}{Ca}, the presence of which
has recently been found to be correlated with the one of \chem{26}{Al} in some
primitive meteorites (Sahijpal et al. 1998). Some weaker evidence has been gathered
about \chem{36}{Cl}, \chem{92}{Nb}, \chem{99}{Tc} and \chem{205}{Pb} (see e.g.
Podosek \& Nichols 1997 for a review and references).
 
The demonstrated existence of short-lived radionuclides in live form in the
early
solar system can usefully constrain the chronology of the nebular and planetary
events at that epoch (e.g. Podosek \& Nichols 1997 for details). From a more
astrophysical
point of view, these observations are generally considered to provide the most
sensitive radiometric probes concerning discrete nucleosynthesis  events that
presumably contaminated the solar system at times between about $10^5$ and
$10^8$ y prior to the isolation of the solar material from the general galactic
material. Of course, this statement assumes implicitly that the radionuclides
of
interest have not been synthesized in the solar system itself. Can such a local
production scenario be rejected right the way ? This question has been
revisited
recently, as briefly mentioned in Sect.~4. 
 
If indeed the short-lived radionuclides that have been present live in the
early
solar system are not of local origin, the message they carry on the chronology
of
the nucleosynthetic events  responsible for a `late pollution' of the solar
system
can obviously not be extracted from the chemical evolution models needed when
one
deals with long-lived chronometers. Instead, a scenario relying on a limited
number
of events has to be constructed. One form of such a `granular' description is
referred to as the `Bing Bang' model
(Reeves 1978, 1979), which envisions the contamination and formation of
the
solar  system in an OB association
during its approximate $10^7$ y lifetime.
\footnote{OB associations are groupings of highly luminous and hot massive
stars of the O and B spectral types. Many OB associations, which are about 30 to 200
parsec accross (1 parsec is 3.26 lightyears), are made of smaller clusters of stars
called OB subgroups containing about 5 to 20 stars, with an average of about 10. One
of the best studied OB associations lies in the Orion cloud complex (e.g. Genzel \&
Stutzki 1989 for a review). OB associations are not bound by gravity. In addition,
their constituting stars explode as supernovae after a rapid evolution not lasting
more than a few tens of millions of years. As a consequence, these associations
dissipate in a few times
$10^7$ y. So, if the solar system had been born in an OB association, this star
grouping would have disappeared a long time ago!}  
A chronology based on these granular
chemical evolution models raises a series of important and difficult questions
related in particular to the type of nucleosynthetic event(s) responsible for
the contamination, the corresponding radionuclide yields, as well as the efficiency
of the pollution. 
 
It has to be emphasized that a Bing Bang type of model does not imply either a
contamination by a single astrophysical source, or a strict relation between
the contamination and the very formation of the solar system. In contrast, the
possibility for a given star to trigger the formation of the solar system and to
contaminate it at the same time has received much attention over the years, and has
recently been scrutinized within hydrodynamical models (e.g. Boss \& Foster 1998,
and references therein).
 
Supernovae, Asymptotic Giant Branch (AGB)  or Wolf-Rayet (WR) stars have
been identified as possible triggering/contaminating agents. They are briefly
reviewed in Sects.~3.2 - 3.4. We also make some specific comments on the short-lived
radionuclides \chem{205}{Pb} (Sect.~3.5) and \chem{146}{Sm} (Sect.~3.6) which have
been explored by Jerry in some of his recent works. 
 
\subsection{Some brief comments on the supernova production of short-lived
radionuclides}

Supernova explosions are spectacular events representing the endpoint of the
evolution of a variety of stars. Exploding massive ($M \gsimeq 10$ M$_\odot$) stars
are classified as Type II supernovae (SNII) if hydrogen lines are present in
their spectra. Some massive stars, and in particular Wolf-Rayet stars (see Sect.~3.4),
may have lost their H envelope at the time of their explosion, which is generally
classified as a Type Ib/c supernova (SNIb/c). Other supernovae, referred to as Type
Ia (SNIa) are associated with the explosion of a compact star, referred to as a
`white dwarf', accreting material from a companion star in a binary system. The
mechanism responsible for the SNIa is completely different from the one leading to
SNII or SNIb/c.

Type II supernovae eject large amounts of
nuclearly processed material into the interstellar medium, and  
could have injected live radionuclides into the solar system forming as a result of
the explosion itself. Much has been written on the subject, and the reader is
referred to e.g. Cameron et al. (1995) for details and further references. We just
want to mention here that Jerry and his collaborators (Waserburg et al.
1998) have proposed a test of the SNII trigger hypothesis based on the
\chem{60}{Fe}/\chem{56}{Fe} ratio that would have to be found in
\chem{26}{Al}-bearing samples if both radionuclides had just been produced by a SNII.
Similar constraints could come from other stable or unstable nuclides that are
predicted to be synthesized in the same SNII zones as the above-mentioned
radionuclides, or in nearby regions. From the available experiments and predicted
SNII yields,  Wasserburg et al. (1998) note that the observed 
\chem{60}{Fe} abundances  appear to be too low
to be compatible with a supernova trigger that injected the \chem{26}{Al} into the
protosolar nebula, the same conclusion also holding for \chem{53}{Mn}.

On the other
hand, it has to be noticed that the SNII contaminating role might be enhanced by
(isotopically anomalous) `shrapnel-like' SN grains which have a higher penetration
efficiency in the forming solar system than the gas from an expanding SN shell
(Margolis 1979). In addition, such grains minimize the danger of having the isotopic
anomalies washed out beyond recognition in the bulk nebular material before the start
of the solar system solidification sequence (of course, this does not exclude grain
vaporisation {\it during} the solidification, which seems to be required by the
analysis of the isotopic anomalies attributed to the radionuclide in situ decay). The
possible role in the SN contamination efficiency of SN `fast moving knots' stressed
by  Arnould \& N{\o}rgaard (1978) may be nicely complementary to the polluting
importance  of grains. There is indeed mounting evidence that fast moving knots are
priviledged locations of grain formation in SN ejecta (Lagage et al. 1996).
Certainly, the details of the contamination are still far from being settled. In any
case, it seems highly plausible that the short-lived radionuclides have been
distributed  heterogeneously in the forming solar system (see also Podosek \& Nichols
1997). This complicated situation may affect quite negatively their chronological
predictive virtues. 

Finally, let us note that other supernova types might also contribute to the
synthesis of short-lived radionuclides. This is especially the case of SNIb/c which
are often interpreted as exploding WR stars. This production (in particular of
\chem{53}{Mn}, \chem{60}{Fe} or \chem{146}{Sm}) could complement the non-explosive WR
contamination (see Sect.~3.4) after a time span that is shorter than the lifetime of
the considered radionuclides.
  
\subsection{Short-lived radionuclides from AGB stars}

After their central hydrogen and helium burning stages,
low- and intermediate-mass stars ($1 \lsimeq M \lsimeq 8$ M$_\odot$) enter the AGB
phase, characterized by red colors and high luminosities. Most AGB stars are observed
to have their surfaces contaminated with ashes of H- and He-burning, including
s-process products.\footnote{Three nucleosynthetic processes are called for in order
to explain the production of the stable nuclides heavier than iron. Two of them rely
on neutron captures. In the so-called s-(slow-)process, a $\beta$-unstable nucleus,
if produced, has in general time to decay before capturing a neutron. In such
conditions, the corresponding nuclear flow is confined to the close vicinity of the
line of nuclear stability, and the s-process thus produces the nuclei lying on this
line.  The reverse situation is encountered in the r-(rapid)process, where the
neutron fluence is so high that the material is pushed deep inside the region of very
neutron-rich
$\beta$-unstable nuclei. If the neutron concentration decreases or even goes to zero,
these nuclei cascade down to the valley of stability until a stable nucleus is
reached. So, the r-process accounts for the neutron-rich stable heavy nuclei, and also
complements the s-process contribution to the nuclei at the bottom of the valley of
nuclear stability which are not bypassed from the r-process flow by a neutron-rich
stable isobar. The third process, referred to as the p-process, accounts for the
stable heavy neutron deficient nuclides. It is made of the destruction of pre-existing
s- or r-nuclides by a variety of nuclear reactions involving the emission of a
nucleon or of an $\alpha$-particle. While the s-process develops during non-explosive
stages of the evolution of a large variety of stars, the r-process is likely
associated with violent stellar phenomena, like supernovae. However, the precise site
for this mechanism has not been identified yet. The p-process probably develops
during SNII, and possibly also during the stages just preceding these explosions
(see e.g. Arnould \& Takahashi 1999 for more details)} Theory relates these
observations to the prediction that the structure of these stars is characterized by
thin H- and He-burning shells, by the recurrent occurence of convectively unstable
so-called `thermal pulses' in the He shell, and by the possibility of transport of
part of the nuclearly processed material to the surface
(`third dredge-up') and into the ISM (by strong winds).
 
Along these lines, it has been suggested by Cameron (e.g. Cameron 1985), and further
studied by Jerry and some collaborators (Wasserburg et al. 1994) that AGB stars
might have contributed appreciably to some short-lived isotopes in the ISM and in the
early solar system. As Jerry is well known to be largely open to scientific
debates about unsettled issues, he is most likely ready to acknowledge that the
astrophysical plausibility of his contamination scenario and its true efficiency
remain to be demonstrated. 
 
Apart from the difficult problem of evaluating in astrophysical terms the
probability
for an AGB star to have been able to contaminate the forming solar system with
short-lived radionuclides at the level required by laboratory data, many
uncertainties remain in the precise evaluation of the radionuclide yields in
terms of
stellar mass and metallicity. One such major uncertainty concerns the
efficiency of
the source of neutrons leading to the development of the s-process in AGB
stars. It
is generally agreed today that \reac{13}{C}{\alpha}{n}{16}{O} is the relevant
neutron producing reaction. The crux of the problem lies elsewhere, and more
specifically in the  prediction from `first principles' of the available amount
of
\chem{13}{C}, which has to exceed largely the one emerging from the CNO burning
of
hydrogen in order to allow a fully developed s-process to operate (this extra
\chem{13}{C} is referred to as `primary'). In these matters, the theoretical
challenge lies in the proper description of the mixing of matter across a H-He
abundance discontinuity that develops in model stars. Such a mixing at the
right
level and with the correct depth distribution is indeed agreed to be a
necessary condition to get the proper s-process.  Recent works on that subject  
consider the effect of rotation (Langer et al. 1999) or of diffusive penetration of
matter from the convective envelope into the underlying radiative layers (e.g. Herwig
et al. 1997). These calculations provide a new
basis for the study of the production of the primary \chem{13}{C}. Still, most
predictions, and in particular those of Wasserburg et al. (1994), view the amount
of that \chem{13}{C} and its variation with position in the star as free parameters.
Related uncertainties could of course be reduced on a case by case basis by trying to
reproduce at best the s-process abundances observed at the surface of a given star
(e.g. Gallino et al. 1998). This procedure would clearly be of the greatest value for
better understanding the \chem{13}{C} production mechanism if quantities like the
stellar masses and metallicities of the considered stars were known, which is often
not the case. 
 
Another serious uncertainty relates to the quantity of C- and
s-process-enriched
matter dredged-up to the surface after each thermal pulse. In this field, too,
significant progress has been made recently (e.g. Mowlavi 1999). The
proper
modeling of the dredge-up process and the reliable evaluation of the
production of primary \chem{13}{C} may actually be closely linked to each
other. They
would open the door to a more consistent study of the yields of short-lived
isotopes
from AGB stars.

Among those radionuclides, \chem{26}{Al} holds a special position. 
It is demonstrated by now that \chem{26}{Al} has been live in the early solar
system (Sect.~3.1), and has also entered it in extinct form, carried by a
variety of
pre-solar grains (Sect.~3.7). As if this were not enough, \chem{26}{Al} also exists
in live form in the present ISM (Sect.~5.1).

On the theoretical side, \chem{26}{Al} is essentially a product of the MgAl
chain of hydrogen burning. This burning can occur in the H-rich shell of AGB stars. It
could also develop in some \chem{12}{C}-rich layers below the H-He composition
discontinuity which might engulf enough protons for the Mg-Al chain to operate
efficiently enough in order to dominate the \reac{26}{Al}{n}{p}{26}{Mg} destruction
due to the neutrons liberated by the radiative burning of the \chem{13}{C} also
produced in these layers. Hot enough pulses for neutrons to be liberated by
\reac{22}{Ne}{\alpha}{n}{25}{Mg} could destroy part at least of the pre-pulse
\chem{26}{Al}. A recent detailed discussion of the net \chem{26}{Al} production by
pulsing AGB stars can be found in Goriely (1999).
These conclusions still
need to be confirmed by AGB models  producing consistently some primary \chem{13}{C}.
Reliable yields from AGB stars are thus still awaiting improved stellar models, even
for the `simple' case of \chem{26}{Al}.

\vskip1truecm
\centerline{EDITOR: FIGURE 2 HERE}
\vskip1truecm

The case of \chem{205}{Pb}, to be considered in more details in Sect.~3.5,
deserves also a brief comment here. The importance of AGB stars as a potential
source of \chem{205}{Pb} has first been envisioned by
Yokoi et al. (1985), and revisited by Jerry and collaborators
(Wasserburg et al. 1994), as well as by Mowlavi et al. (1998). The latter authors
performed
the first quantitative study of the ability of \chem{205}{Pb} to {\it survive}
in
neutron-free locations in between thermal pulses. It is concluded by
Mowlavi et al. (1998) that the chances for a significant
\chem{205}{Pb} yield from AGB stars are likely to increase with the stellar
mass for
a given metallicity, or to increase with decreasing metallicity for a given
stellar
mass. However, again, quantitative predictions of \chem{205}{Pb} yields from
AGB 
stars are lacking due to the failure of modeling consistently the s-process in
those
stars.

\subsection{WR stars: short-lived radionuclide contaminators of the early solar
system ?}

Wolf-Rayet (WR) stars are generally considered to be the normal evolutionary phase
of very massive stars ($M \gsimeq 20$ to 80 M$_\odot$, depending upon their initial
composition). Their structure, evolution and surface compositions are governed by
huge non-explosive mass losses which can amount to as much as $10^{-4}$ M$_\odot$/y.
As a result of these spectacular winds, products of central hydrogen and
even of central He burning can appear at the stellar surface, and be ejected into the
interstellar medium. Apart from their role of strong interstellar medium chemical
contamination, these winds have also a substantial dynamical effect of the WR
surroundings. These stars are predicted to explode as SNIb/c (see above). The reader
is referred to e.g. Arnould et al. (1997a) for a more detailed review of the observed
properties of WR stars, as well as of some theoretical considerations.

Much progress has been made recently in the modeling of these stars. Many important
observed features, like their luminosities, surface chemical compositions, or
statistics in regions of constant star formation rate, as well as in starburst
locations, can now be nicely accounted for by single non-rotating model stars (e.g.
Maeder 1991, Maeder \& Meynet 1994). In spite of this success, uncertainties of course
remain in the models. However, it has to be stressed also that the impact on the
predictions of these uncertainties is largely reduced under the constraint that the
largest possible body of observed intrinsic WR properties have to be reproduced at
best.\footnote{An anonymous referee correctly argues that due account of observational
constraints helps reducing uncertainties for all types of models, and not only in the
WR case. However, it has to be emphasized that the difference between the number of
theoretical `degrees of freedom' and the observational constraints is much smaller
in the WR case than in all the other possible radionuclide producers (like AGB stars,
novae or supernovae of various types)} 
This favorable situation, combined with the fact that the modeling of single
non-rotating WR stars is {\it immensely} simpler than the one of all the other
short-lived radionuclide producers proposed up to now (and in particular of the
AGB
stars; see Sect.~3.3), forces to conclude that {\it the predicted WR yields
are
likely to reach a level of reliability that cannot be obtained in the other
cases}. Of
course, one has to remain alert to the fact that the WR structure and
nucleosynthetic
yield predictions might be affected by rotation, as well as by the binary
nature of a
non-negligible fraction of the WR stars. These effects are briefly discussed below.
 
The production of short-lived radionuclides of cosmochemical
interest has been calculated in the framework of 
detailed evolution models for a large variety of non-rotating WR stars with
different
initial masses and compositions, and with the use of extended nuclear reaction
networks. We first briefly review the special case of \chem{26}{Al} before
considering
the yields of other radionuclides of relevance.

\subsubsection{The \chem{26}{Al} production by non-rotating WR stars}
 
A detailed discussion of the production of \chem{26}{Al} by the MgAl chain of
hydrogen burning developing in WR stars has been conducted recently by
Arnould et al. (1997b)
and Meynet et al. (1997). Figure 3 displays some of their results in the form of the
total
mass $M_{\rm 26}(M_{\rm i},Z)$ of
\chem{26}{Al} ejected by WR stars with various initial masses $M_{\rm i}$ and
metallicities $Z$. It shows that the
\chem{26}{Al} yields increase with initial mass and $Z$, the $Z$ dependence
being
approximated by $M_{\rm 26}(M_{\rm i},Z) = (Z/Z_\odot)^2M_{\rm 26}(M_{\rm
i},Z_\odot)$, where $Z_\odot = 0.02$ is the solar metallicity. It is worth
noticing
that the $^{26}$Al yields of Meynet et al. (1997) are in qualitative agreement with
those of Langer et al. (1995), even if these two sets of models greatly
differ in their physical ingredients. This illustrates the statement made above
concerning the reduced impact of remaining uncertainties in the models once
they are
constrained by observation.
 
\vskip1truecm
\centerline{EDITOR: FIGURE 3 HERE}
\vskip1truecm

\subsubsection{Effect of rotation on the WR production of\, \chem{26}{Al}}

Rotation induces numerous dynamical instabilities in stellar interiors, and in
particular meridional circulation and shear
turbulence (e.g. Zahn 1992). The related mixing of the chemical species can
deeply modify the chemical structure of a star and its evolution, and may in
particular have important consequences on the $^{26}$Al
production by WR stars. At present, very few rotating WR models address this
question
in detail (see Langer et al. 1995 for some preliminary results), so that  
it is certainly premature to quantitatively assess the possible role rotation
plays in
that respect. However, it seems safe to say that rotation increases the quantity
of
$^{26}$Al ejected by WR stellar winds. This claim is made plausible by Fig. 4,
which
shows the structural evolution during their H- and He-burning phases of two 60
M$_\odot$ stars that just differ by their rotational velocities. It appears
that 

\noindent 1) The size of the convective core is enhanced by rotation;

\noindent 2) Rapidly rotating stars may enter the WR phase while still on the
Main
Sequence. Moreover the surface abundances characteristic
of their WNL and WC phases are not due to the mass loss
which uncovers core layers, but result from diffusive mixing in the
radiative zones. As a consequence, the evolution of the surface abundances are
much
smoother in rotating models;

\noindent 3) The WR lifetime increases with $\Omega/\Omega_c$. In
particular the WN stage during which $^{26}$Al is ejected, as well as the WN/WC
transition phase, become much longer, so that much more mass is
ejected during these phases. As a numerical example, the non-rotating model
sketched in Fig. 4 ejects about 5 M$_\odot$ during the WN phase, while about 50
M$_\odot$ is ejected by the rotating model during the same phase.
 
\vskip1truecm
\centerline{EDITOR: FIGURE 4 HERE}
\vskip1truecm
 
The inference that rotation can enhance the \chem{26}{Al} yields of WR stars is
confirmed by the numerical simulations of Langer et al. (1995) (see also
the presupernovae evolution models for rotating massive stars computed by
Heger 1998). Rotation also lowers the
minimum initial mass of single stars which can go through a WR stage with a
concomitant ejection of \chem{26}{Al}. It is thus likely to increase the net
amount
of \chem{26}{Al} ejected by WR stars in the ISM. This may be of relevance both
to
cosmochemistry and to $\gamma$-ray line astrophysics (Sect.~5) as well.

\subsubsection{Effect of binarity on the WR production of \chem{26}{Al}}

Tidal interactions in close binary systems may considerably modify the
evolution of the two stellar components with respect to the one they would
experience as isolated stars. Mass transfer by
Roche Lobe Overflow can act as a strong stellar wind, and thus reduce the
critical
initial mass for producing single WR stars. Before any mass transfer, tidal
effects
are also expected to deform the star and therefore induce instabilities
reminiscent of those induced by rotation. To our knowledge, the latter effect
has
never been studied in any detail, even if it might have important consequences,
like
the homogenization of the stars, and the related inhibition of 
mass transfer. Other effects remain to be explored, like the impact of
colliding
winds on the mass transfer process, or even the very nature of the grains which
can condense in colliding winds
(see Sect.~3.7). Thus the effect
of binarity on the WR star formation and evolution, and on the corresponding
$^{26}$Al production cannot be evaluated with confidence at this time. Some
preliminary estimates (Braun \& Langer 1995) lead to the conclusion that `only for
stars
with masses
$\le$ 40 M$_\odot$, binarity has the potential to increase the $^{26}$Al yield
compared to the single star case' (see also Langer et al. 1998). The fact that the
situation may be quite different in more massive stars can be interpreted in
the
following way: after their Main Sequence, single high mass stars go through a
LBV
phase characterized by especially high stellar winds (see Fig.~4). Additional mass
losses triggered by binarity would thus not induce more than a perturbation in the
evolution
of these stars. In contrast, lower initial mass stars not suffering the LBV
winds
would be drastically affected by Roche Lobe Overflow mass transfer, which acts
as a
strong wind not operating if the corresponding stars were isolated.
 
The extent in the reduction of the \chem{26}{Al} that can be ejected in the
ISM by
$M > 40$ M$_\odot$ WR members of binary systems cannot be assertained at this
point.
It has clearly to depend on the fraction of the mass lost by the WR star which
is
accreted by its companion, and thus withdrawn from the ISM. In this
scenario, however, one may wonder about the fate of the accreting companion,
and
about its
net production or destruction of \chem{26}{Al} resulting from its evolution.
Clearly, the effect of binarity on the net \chem{26}{Al} outcome by WR stars
largely remains to be studied.

\subsubsection{The WR production of other short-lived radionuclides}
 
\vskip1truecm
\centerline{EDITOR: FIGURE 5 HERE}
\vskip1truecm

A complementary detailed study of the production by non-rotating WR stars of
other
short-lived radionuclides of astrophysical and cosmochemical interest has been
conducted by 
Arnould et al. (1997b). In short, their main results may be summarized as follows:
  
%\begin{list}{(\arabic{obj})}{\usecounter{obj}\setlength{\leftmargin}{0pt}
%              \setlength{\labelwidth}{0pt}\setlength{\itemindent}{5pt}
%              \setlength{\listparindent}{\parindent}
%              \setlength{\parsep}{\parskip}
%              }
%\vskip-2truemm
%\item

\noindent (1) The neutrons released by \reac{22}{Ne}{\alpha}{n}{25}{Mg} during
the
He-burning phase of the considered stars are responsible for a s-type
process leading to the production of a variety of $A > 30$
radionuclides. In the absence of any chemical fractionation between the
relevant 
elements, it is demonstrated that
\chem{36}{Cl}, \chem{41}{Ca} and \chem{107}{Pd} can be produced by this
s-process in
a variety of WR stars of the WC subtype (see Fig.~4) with different initial masses and
compositions
at a {\em relative} level compatible with the meteoritic observations. For a
$60$ \ms $\,$ star with solar metallicity, Fig.~5 shows that this
agreement can be obtained for a time $\Delta^\ast \approx 2\times10^5$ y, where
$\Delta^\ast$ designates the time elapsed between the last
astrophysical event(s) able to affect the composition of the solar nebula and
the solidification of some of its material (e.g. Wasserburg 1985). More
details concerning other model stars are given by Arnould et al. (1997b);

%\item 
\noindent (2) To the above list of radionuclides, one of course has to add
\chem{26}{Al} (see above). The canonical
value $(\chim{26}{Al}/\chim{27}{Al})_0 = 5\times10^{-5}$ (MacPherson et al.
1995),
while
not reached in the 60 \ms $\,$ star displayed in Fig.~5, can be obtained from
the
winds of $M \geq 60$ \ms $\,$ stars with $Z > Z_\odot$ under the same type of
assumptions as the ones adopted to construct Fig.~5. Let us also note that the
WR
models can account for the correlation  between \chem{26}{Al} and \chem{41}{Ca}
observed in some meteorites  (Sahijpal et al. 1998);

%\item 
\noindent (3) Too little \chem{60}{Fe} is synthesized;
 
%\item 
\noindent (4) An amount of \chem{205}{Pb} that exceeds largely the experimental
upper
limit set by Huey \& Kohman (1972), but which is quite compatible with the value
reported by
Chen \& Wasserburg (1987), is obtained not only for the model star displayed in Fig.~5, but
also
for the other cases considered by
Arnould et al. (1997b). The interesting case of \chem{205}{Pb} is discussed further
in Sect.~3.5;

%\item
\noindent (5) More or less large amounts of \chem{93}{Zr}, \chem{97}{Tc},
\chem{99}{Tc} and \chem{135}{Cs} can also be produced in several cases, but
these predictions cannot be tested at this time  due to the lack of reliable
observations.
%\end{list}

It has to be remarked that the above conclusions are derived without taking
into
account the possible contribution from the material ejected by the eventual SNIb/c
explosion of the considered WR stars.
This SN might add its share of radionuclides that are not produced
abundantly enough prior to the explosion. This concerns in particular
\chem{53}{Mn}, \chem{60}{Fe} or \chem{146}{Sm}, the latter case being discussed
further in Sect.~3.6. One has also to acknowledge that the above
conclusions
sweep
completely under the rug the possible role of rotation and binarity in the WR
yields.  

From the results reported above, one can try estimating if indeed there is
any chance for the contamination of the protosolar nebula with isotopically
anomalous WR wind material at an {\em absolute} level compatible with the
observations. In the framework of Fig.~5, this translates into the possibility
of obtaining reasonable dilution factors $d(\Delta^\ast)$. A qualitative
discussion of this  highly complex question based on a quite simplistic
scenario is presented by Arnould et al. (1997b). In brief, it is concluded that
astrophysically plausible situations may be found in which one or several WR
stars with masses and metallicities in a broad range of values could indeed
account for some now extinct radionuclides that have been injected live into
the forming solar system (either in the form of gas or grains). Of course, a
more definitive conclusion would have to await the results of a more detailed
model that takes into account the high complexity of the WR circumstellar
shells, and of their interaction with their surroundings, demonstrated by
observation and suggested by numerical simulations. Concomitantly, the
possible role of WR stars, either isolated or in OB associations, as triggers
of the formation of some stars, and especially of low-mass stars, should be
scrutinized.

\subsection{\chem{\rm 205}{\rm Pb}: a short-lived s-process chronometer?}
\label{Sect:Pb205}

Among the short-lived radionuclides of potential cosmochemical and
astrophysical
interest, \chem{205}{Pb} ($t_{1/2} = 1.5 \times 10^7$ y) has the distinctive
property
of being of pure s-process nature, at least if the \chem{204}{Tl} $\beta$-decay
competes successfully with its neutron capture in stellar interiors. This
remarkable
feature has motivated the study of the chronometric virtues of the
\chem{205}{Pb} - \chem{205}{Tl} pair (Blake et al. 1973, Blake \& Schramm 1975). 

This early work has led its authors to express some
doubts about the possibility to rank \chem{205}{Pb} as a reliable s-process
clock. Apart from the fact that the level of the possible \chem{205}{Pb}
contamination of the early solar system was very poorly known [only an upper
limit of about $9 \times 10^{-5}$ being available for the
(\chem{205}{Pb}/\chem{204}{Pb})$_0$ ratio (Huey \& Kohman 1972)], this pessimism
related to
the realization that electron captures by the thermally populated 2.3 keV first
excited state of
\chem{205}{Pb} might reduce drastically the \chem{205}{Pb} effective lifetime
in a
wide range of astrophysical conditions. Of course, the likelihood of a late
injection of \chem{205}{Pb} into the (proto-)solar nebula was reduced
accordingly.

This conclusion has been demonstrated to be invalid, in certain s-process
conditions
at least. The \chem{205}{Pb} destruction into \chem{205}{Tl} by electron
captures may
indeed be efficiently hindered by the reverse transformation, which is made
possible
as a result of the \chem{205}{Tl} bound-state $\beta$-decay. The nuclear
aspects of
this question have been analyzed in considerable detail by Yokoi et al. (1985), who
have
shown on grounds of schematic astrophysical models that the possible level of
\chem{205}{Pb} s-process production may be large enough to justify a renewed
interest
for the \chem{205}{Pb} -- \chem{205}{Tl} pair. 

The work of Yokoi et al. (1985) has indeed triggered further observational,
experimental and theoretical efforts. In particular, a new measurement of the
(\chem{205}{Pb}/\chem{204}{Pb})$_0$ ratio has been attempted by Jerry and one
of his
co-workers (Chen \& Wasserburg 1987), leading to a value of about $3 \times 10^{-4}$. On
the
other hand, some experiments are currently deviced in order to obtain a direct
measurement of the \chem{205}{Tl} -- \chem{205}{Pb} mass difference with high
precision (Vanhorenbeeck 1998). This quantity, which is still somewhat
uncertain,
affects quite drastically the predicted
\chem{205}{Tl} bound-state $\beta$-decay. Finally, more reliable estimates of
the
\chem{205}{Pb} yields have been obtained through detailed s-process
calculations
performed with the help of realistic model stars. This concerns in particular
Wolf-Rayet stars (see Sect.~3.4). In view of the rather high predicted
\chem{205}{Pb} WR yields (Fig.~5), one might wonder if the availability of
reliable laboratory data for (\chem{205}{Pb}/\chem{204}{Pb})$_0$ could not help
discriminating a WR origin from  other potential \chem{205}{Pb} sources, such
as 
AGB stars (see Sect.~3.3). Of course, this question would be most profitably
addressed if the \chem{205}{Pb} AGB yield predictions could be put on a safe
footing.
 
In short, one may conclude from the above considerations that much
cosmochemical,
nuclear and astrophysics work remains to be done for giving a chance to
\chem{205}{Pb} to gain the status of a reliable short-lived s-process
chronometer. 

\subsection{\chem{\rm 146}{\rm Sm}: a short-lived p-process radionuclide}
\label{Sect:Sm146}

There is now strong observational evidence for the existence in the early solar
system of the two p-process radionuclides 
$^{92}{\rm Nb^g}\, (t_{1/2} = 3.6\, 10^7$ y) and 
$^{146}{\rm Sm}\, (t_{1/2} = 1.03\, 10^8$ y) (Harper 1996, and references
therein). The case of $^{92}{\rm Nb^g}$ has been discussed by Rayet et al. (1995),
who
conclude that various uncertainties in the level of production of this
radionuclide
make rather unreliable at this time the development of a \chem{92}{Nb}-based
p-process chronometry.

As far as \chem{146}{Sm} is concerned, the study of its potential as a
p-process
chronometer has been pioneered by  
Audouze \& Schramm (1972). This work has triggered a series of meteoritic,
nuclear physics and astrophysics investigations, which have helped clarify
many
aspects of the question. In particular, Jerry and his collaborators
(Prinzhofer et al. 1989) have contributed in an essential way to the reduction of
the early
uncertainties on the amount of
\chem{146}{Sm} that has been injected live into the solar system through high
quality
measurements of the
\chem{142}{Nd} excess observed in certain meteorites as the result of the in
situ
\chem{146}{Sm} $\alpha$-decay.  
More specifically, it is concluded to-day that
(\chem{146}{Sm}/\chem{144}{Sm})$_0 =
0.008 \pm 0.001$,
\chem{144}{Sm} being the stable Sm p-isotope. One can attempt building up a
p-process 
chronometry on this value if the corresponding isotopic production ratio can be
estimated reliably enough at the p-process site.

Much work has been devoted to the modeling of the p-process in massive stars,
and
especially in SNII (e.g. Arnould et al. 1998). In spite of this, the 
production ratio $P \equiv$ \chem{146}{Sm}/\chem{144}{Sm} remains quite
uncertain,
being estimated by Somorjai et al. (1998) to lie in the $0.7 < P < 2$ range in
(spherically
symmetric) SNII models. This unfortunate situation relates in
part to astrophysical problems, and in part to nuclear physics uncertainties,
especially in the \reac{148}{Gd}{\gamma}{\alpha}{144}{Sm} to
\reac{148}{Gd}{\gamma}{n}{147}{Gd} branching ratio (e.g. Rayet \& Arnould 1992), even
if the
prediction of this ratio has recently gained increased reliability. This
improvement comes from the direct measurement of the
\reac{144}{Sm}{\alpha}{\gamma}{148}{Gd} cross section down to energies very
close to those of direct astrophysical interest, complemented with a better nuclear
reaction model concerning in particular the appropriate low-energy $\alpha$-particle
nucleus optical potential (Somorjai et al. 1998).

The resulting astrophysical rate is
predicted to be 5 to 10 times lower than previous estimates in the temperature range
of relevance for the
production of the Sm p-isotopes.  By application of the detailed balance
theorem, the
rate of the reverse \reac{148}{Gd}{\gamma}{\alpha}{144}{Sm} of direct
astrophysical
interest is reduced accordingly. This implies a lowering of the \chem{144}{Sm}
production, and favors concomitantly the \chem{146}{Sm} synthesis through the
main
production channel
\chem{148}{Gd}($\gamma$,n)\chem{147}{Gd}($\gamma$,n)\chem{146}{Gd}($\beta^+$)\chem{146}{Sm}.
The net effect of the revised
\reac{148}{Gd}{\gamma}{\alpha}{144}{Sm} rate is thus an increase of the $P$
values.

Other nuclear problems add to the uncertainty in the evaluation of $P$. As
noted
above, this concerns in particular the \reac{148}{Gd}{\gamma}{n}{147}{Gd}
reaction,
for which no experimental information can be foreseen in a very near future in
view
of the unstable nature of \chem{147}{Gd} ($t_{1/2} \approx 38$ h). An analysis
of the
sensitivity of $P$ to this rate has been conducted by Rayet \& Arnould (1992) for
SN1987A. 

In view of the difficulty of predicting $P$ reliably in SNII, one has to
conclude that 
\chem{146}{Sm} cannot be viewed at this point as a reliable p-process
chronometer.
Further uncertainties arise, relating in particular to the possibility of
additional, and still poorly studied,
\chem{146}{Sm} production by SNIa, as well as by exploding SNIb/Ic  
WR stars, which could in fact release the same suite of
radionuclides as SNII. 

\subsection{Extinct short-lived radionuclides in the solar system}

The year 1987 has marked the start of the discovery and isolation
of meteoritic grains that are interpreted as specks of stardust having survived
the formation of the solar system. Since that time, a remarkable series of
dedicated laboratory work has been conducted, and has led by now to the
identification and analysis of
a long suite of such presolar grains. They are made of refractory materials of various
types (diamond, SiC, graphite, corrundum, silicon nitride). Tiny subgrains, in
particular Ti-, Zr- and Mo-carbide or TiC subgrains, have even been found in
graphite or SiC grains, respectively (see Bernatowicz \& Zinner 1997 for many
contributions on presolar grains). All the analyzed elements contained in these
grains exhibit much larger anomalies than those found  in the  material that
condensed in the solar system itself. This is interpreted as the largely
undiluted
nucleosynthetic  signature of specific stellar sources.

This rule applies in particular to the \chem{26}{Mg} excesses attributed to
the in situ decay of \chem{26}{Al} observed in presolar silicon carbide,
graphite and
oxide grains, as demonstrated by e.g. Fig.~14 of MacPherson et al. (1995) (see also
the
reviews on specific grain types in Bernatowicz \& Zinner 1997).
The initial \chem{26}{Al}/\chem{27}{Al} ratio inferred to have been present in
the
analyzed grains vary from about $10^{-5}$ to values as high as about 0.5, to be
compared to the canonical solar system value (\chem{26}{Al}/\chem{27}{Al})$_0
\approx 5 \times 10^{-5}$ (Sect.~3.1). The highest reported ratios obviously
put
particularly drastic constraints on the \chem{26}{Al} production models,
especially
when the Al data are complemented with correlated isotopic anomalies in other
elements, and in particular in C, N, and O. As discussed in some detail by
Arnould et al. (1997a),  WR stars could well explain even the highest reported
\chem{26}{Al}/\chem{27}{Al} ratios, but might have some problem accounting for
the
isotopic composition of, in particular, nitrogen. In addition, one has to
acknowledge that there is no clear indication yet that the types of
grains loaded with large \chem{26}{Al} amounts can indeed condense from the WR
winds\footnote{As emphasized in several places (Arnould et al. 1997a, 1997b), binarity
might affect the chemistry and nature of the grains known to form around WR
stars. In
fact, it has been speculated that a recently discovered singular presolar SiC
grain
might find its origin in a WR-O binary system (Amari et al. 1999)}. 
 
Another example is provided by an extraordinary neon component, referred to
as Ne-E(L), which is carried by presolar graphite grains, and is made of almost
pure
\chem{22}{Ne} (e.g. Amari et al. 1995). This remarkable feature is generally
interpreted in terms of the in situ decay of the ultra-short radionuclide
\chem{22}{Na} ($t_{1/2}
\approx 2.6$ y). In view of its short lifetime, the production of \chem{22}{Na}
in the
thermonuclear framework requires the consideration of explosive situations. The
first
explicit connection of this sort has been made by Arnould \& Beelen (1974) through
detailed
explosive H burning calculations. They substantiated the later view
(Clayton \& Hoyle 1976) that Ne-E is hosted by nova grains.\footnote{The two forms of
Ne-E 
identified today, Ne-E(L)
and Ne-E(H), were not known at that time. It is generally considered by now
that the less extreme \chem{22}{Ne} enrichments exhibited by Ne-E(H) do not
require
a \chem{22}{Na}-decay origin. In fact, the ejecta of AGB and WR stars are
enriched
in \chem{22}{Ne}. it may also be of interest to note that WR stars have been
considered as responsible for a \chem{22}{Ne} excess observed in the Galactic
Cosmic
Rays (Arnould 1984)}  Over the years, many calculations have been carried
out
along these lines (e.g.  Jos\'e et al. 1998 for a recent
study). Supernovae could also be
\chem{22}{Na} producers through explosive C burning, as demonstrated by the
early
calculations of Arnett \& Wefel (1978), and confirmed by more recent studies (e.g.
Woosley \& Weaver 1995).

Some graphite grains also
carry \chem{41}{K} excesses of up to two times solar that are attributed to the
in situ decay of
\chem{41}{Ca} ($t_{1/2} = 10^5$ y). From these observations, the intial
\chem{41}{Ca}
abundances are inferred to lie in the $10^{-3} \lsimeq
\chim{41}{Ca}/\chim{40}{Ca}
\lsimeq 10^{-2}$ range
(Amari \& Zinner 1997). The \chem{41}{Ca} production may be due to a s-process-type of
neutron
captures associated with He burning in AGB (Wasserburg et al. 1995) or in WR
(Arnould et al. 1997a) stars. However, the (uncertain)
\chem{41}{Ca} load of the AGB winds is predicted to be too low to
account for the observations. The situation is slightly more favorable in the
case of
the WR stars, even if the highest observed ratios remain out of reach.
Supernovae can
also eject some
\chem{41}{Ca} whose abundance relative to \chem{40}{Ca} can be of the order of
$10^{-2}$ in a variety of O- and C-rich layers
(Woosley \& Weaver 1995). A suite of isotopic anomalies accompany the
\chem{41}{K} excess, and in particular an inferred \chem{26}{Al}/\chem{27}{Al}
ratio
ranging typically between 0.01 and 0.1.
 
In order to account for these correlated anomalies, large scale mixing of 
ad hoc amounts of SNII layers with different compositions has been proposed
(Travaglio et al. 1998). If the very existence of such a mixing is supported by
observation (e.g. Nagataki 1999 for references), it remains to model it and
the associated composition pattern in a reliable way, which is far from being the
case at present (e.g. Nagataki 1999 for comments and references). Even
the presupernova composition may be less well predictable than generally imagined
(e.g. Bazan \& Arnett 1998). In view of these many difficult pending problems, it is
quite risky to try evaluating the plausibility of
the blends of supernova layers that are called for in order to fit the observed
anomalies. 
  
Finally, evidence for the presence of \chem{44}{Ti} ($t_{1/2} \approx 60$ y) in
some
graphite and SiC-X grains is provided by \chem{44}{Ca} excesses that translate
into
\chem{44}{Ca}/\chem{40}{Ca} ratios of up to about 140 times solar
(Amari \& Zinner 1997).
This radionuclide is predicted to be produced in the so-called $\alpha$-rich
freeze-out developing in the layers of a SNII located just outside the
forming neutron star. The corresponding yields are thus extremely sensitive to
the
still uncertain details of the physics of the explosion. The isotopic anomalies that
correlate with the \chem{44}{Ca} excess are often considered to demonstrate a SNII
origin of the carrier grains (Nittler et al. 1996). Again, a large mixing of ad hoc
amounts of different SN layers has to be postulated, and the possibility  of
getting just the right mixing of course remains to be demonstrated.  
 
It has also been proposed (Clayton et al. 1997) that the correlated anomalies carried
by the SiC~-~X grains, including the \chem{44}{Ti}-related anomaly, present evidence
for a SNIa origin. Clearly, this model alleviates the mixing problem just mentioned,
but raises other highly complex questions. In particular, and as acknowledged by
Clayton et al. (1997), the progenitors and the evolution to explosion of SNIa are far
less well understood than in the SNII case. In such conditions, it is not known at
this time if nature can really provide the conditions that would be required for
producing SiC-X grains with the right load of isotopic anomalies.
 
\section{Spallative production of short-lived radionuclides}

 Clearly, a large variety of radionuclides
are produced continuously in the present extraterrestrial solar system matter,
as
well as in terrestrial samples, by spallation reactions induced by Galactic
Cosmic
Rays (GCRs) or by Solar Energetic Particles (SEPs) (e.g.
Michel 1998). However, such a production cannot account for the abundances
of
extinct radionuclides derived from the observed isotopic anomalies mentioned
above,
at least if the spallation processes have operated in the early solar system at
a
level commensurable with their present efficiency. The only hope for a local
production to be viable is thus to call for an enhanced spallation production
which
could be associated with the increased SEP production of the young Sun,
especially
in its T-Tauri phase. The plausibility of such an enhancement is supported by a
variety of observations of `Young Stellar Objects (YSOs)'. 

Jerry and one of the authors (M.A.) have revisited this question more than a
decade
ago (Wasserburg \& Arnould 1987) in an attempt to identify a possible relationship
between the
early solar activity and the extinct \chem{26}{Al} and \chem{53}{Mn} in
meteorites.
This is just one in a series of investigations dealing with the spallation
scenario,
which has been the subject of a recent and renewed interest. One has to
aknowledge
that, in spite of some success, the various studies
conducted along these lines reach the conclusion that it appears difficult to
account
for the production of all the relevant short-lived radionuclides in proportions
compatible with the observations, at least without invoking a prohibitive
number of
`tooth fairies' (e.g. Podosek \& Nichols 1997 for references; also
Lee et al. 1998 and Sahijpal et al. 1998).

It may be worth noting that Wasserburg \& Arnould (1987) stress the possible interest,
and identify some consequences, of envisioning spallation scenarios in
which the ratio of $\alpha$-particle to proton fluences would be higher than in
the
solar nebula. Such a situation might be prevaling around H-depleted WR stars.
In
this scenario, spallation and thermonuclear production of short-lived
radionuclides
might just be indissolubly complementary. It is our opinion that these
speculations deserve further scrutiny.

\section{Gamma-ray line astrophysics}

As reviewed in the previous sections, short- or ultra-short-lived radionuclides
have left their signatures in solar system solids or meteoritic presolar
grains in the form of excesses of their daughter products. They may also be
identifiable in the present ISM if their decay leads to a
substantial feeding of a nuclear excited state of the daughter products. In
such a
situation, their electromagnetic de-excitation  produces $\gamma$-ray lines
with specific energies, usually in the MeV domain.

Gamma-ray astrophysics complements in a very important way the study of
extinct radionuclides in meteorites. In particular, it provides information on
the
present-day production of these nuclides; it also allows in some instances a
direct identification of their nucleosynthetic sources, while the
cosmochemical inferences are  necessarily indirect. This
complementarity has not escaped  Jerry's curiosity (Qian et al. 1998).

%Clayton and his collaborators pioneered
%the study of the detectability of \ga-ray lines from the decay of radioactive nuclei
%synthesized in supernova explosions, about  30 years ago.
%This research was in fact the natural follow-up of the predictions
%(Bodansky et al. 1968) that \chem{56}{Fe} is produced in a supernova as the
%radioactive \chem{56}{Ni}, its \chem{56}{Co} decay product ($t_{1/2} \approx 77$
%days) being in its turn responsible for the powering of the optical supernova light
%curves.
%This early work has been followed by the prediction that additional radionuclides
%could be detectable as $\gamma$-ray line emitters (e.g. \cite{Clayton82}). With time,
%it has also been realized that supernovae are  not the sole objects
%of relevance in this field, and that various radionuclides could also be ejected from
%novae, or even mass-losing AGB or WR stars.

On the observational side, $\gamma$-ray line astronomy has received
considerable
momentum in the 80s with the detection of \chem{26}{Al} in the Milky Way. It
has
now been promoted to a mature astrophysical discipline following the discovery
of
additional lines from the \chem{56}{Ni} $\longrightarrow$
\chem{56}{Co} $\longrightarrow$
\chem{56}{Fe} disintegration chain in the supernovae SN1987A and SN1991T, from
the
\chem{57}{Co} $\longrightarrow$ \chem{57}{Fe} in SN1987A, and from
\chem{44}{Ti}
$\longrightarrow$ \chem{44}{Sc} $\longrightarrow$ \chem{44}{Ca} in the CasA and
J052-4642 (Iyudin et al. 1998) supernova remnants.
 
These observations have been complemented with a substantial amount of
theoretical effort in order not only to interpret the available data, but also
to
predict other potential $\gamma$-ray line candidates. Among them, \chem{22}{Na}
and
\chem{60}{Fe} are of special interest (see 
Arnould \& Prantzos 1999 or Diehl \& Timmes 1998 for recent reviews). 
  
The production  of a radionuclide at a high enough level and its
decay to an excited state of its daughter nucleus are necessary but not
sufficient conditions for it to be an interesting candidate for $\gamma$-ray
astronomy. Other factors indeed
play a key role. This concerns in particular the decay
lifetimes, which enter the problem through the fact that the
  production of the nuclei of interest  takes place in
environments initially opaque to $\gamma$-rays; these photons are thus degraded
in
energy as they interact with the  surrounding material. The
$\gamma$-ray lines have in fact a significant probability to be detectable only
if the
matter densities, and thus the opacities, become low enough on
timescales shorter than the radioactive decay lifetimes. These conditions can
be met
in AGB or WR stars, which eject through extensive steady winds a substantial
fraction of their relatively low-density outer material that can be enriched
with
certain $\gamma$-ray line candidates (see Sect.~5.1 for a brief account of the
\chem{26}{Al} production in relation with $\gamma$-ray line observations).
However,
most radionuclides of interest are produced in explosive events of the nova or
supernova types. In the latter case, the synthesis of the nuclides of
relevance
takes place in highly opaque deep layers, so that especially drastic
constraints are
put on the
$\gamma$-ray line observability. The situation is, in fact, quite different
when
dealing with SNIa explosions of low-mass stars, or
with SNII explosions of massive stars. The SNIa ejecta
reach low opacites much more quickly than the SNII ones in view of their lower
masses (about 1 \ms) and larger ejection velocities (in excess of $1.5 \times
10^4$
km/s). More specifically, the ejecta become transparent to
$\gamma$-ray photons  typically  after a few weeks in SNIa and only
after about one year in the SNII case. Even if SNIa provide better detection
conditions, they forbid in particular the observation of the $\gamma$-ray line
associated with the decay of the  important nuclide \chem{56}{Ni}, whose
half-life is
only 6 days.

\subsection{The case of \chem{26}{Al}}

Not content with playing a very special role in cosmochemistry, \chem{26}{Al}
has
also been responsible for the birth of 
$\gamma$-ray line astrophysics with the observation in the present ISM of the
1.8
MeV $\gamma$-ray line emitted following the de-excitation of the \chem{26}{Mg}
produced by the \chem{26}{Al} $\beta$-decay.

The data available to-date indicate that the present ISM contains
about 2 M$_\odot$ of
\chem{26}{Al}, the distribution of which excludes (i) a unique point source in
the
galactic center, (ii) a strong contribution  from the old stellar population of the
galactic bulge, and (iii) any class of sources involving a large number of
sites with
low individual yields, like novae or low-mass AGB stars. In contrast, they
favor
massive stars (WR stars and/or SNIb/Ic and SNII) as the
\chem{26}{Al} production sites (e.g. Kn\"odlseder et al. 1999). They suggest in
fact that most of the
\chem{26}{Al} is made by high metallicity WR stars in the inner Galaxy.

The \chem{26}{Al} yields from non-rotating WR stars predicted as reviewed in
Sect.~3.4 have been used to evaluate quantitatively the virtues of these
stars as
sources of the 1.8 MeV line in the present ISM. Figure 6 shows the mass of live
$^{26}$Al deposited by the winds of WR stars  in rings of increasing
galactocentric
radius. This estimate is based on the metallicity dependent yields computed by
Meynet et al. (1997), and on their assumptions concerning in particular the Initial
Mass
Function, and the galactocentric radius dependence of the star formation rate
and of
the metallicity (for details, see their Sect.~4.1).
The signature of
the
$\sim 5$ kpc ring of molecular clouds located at the galactocentric radius of
about 5
kpc is clearly seen. It is also predicted that more than half of the total
\chem{26}{Al} mass is contained within this ring. 

The integration of the histogram of Fig. 6 over the galactic radius leads to a
total
galactic mass of 1.15 M$_\odot$ produced solely by non rotating non-exploding
WR
stars. Due consideration of various uncertainties leads to masses in the
probable 
0.4-1.3 M$_\odot$ range (Meynet et al. 1997), so that the considered stars might
account
for 20 to 70\% of the present galactic \chem{26}{Al}. 
 
\vskip1truecm
\centerline{EDITOR: FIGURE 6 HERE}
\vskip1truecm

As already stressed in Sect.~3.4, these results may have to be
changed
when rotation and/or binarity are taken into account. In fact, rotation
increases the
total contribution of WR stars to the galactic
\chem{26}{Al} by lowering the minimum initial mass star of single stars which
can go
through a WR phase. It is not possible to be more quantitative on this point at
present. A detailed evaluation of the global
\chem{26}{Al} galactic input by rotating WR stars indeed requires the
computation
of a grid of models with different initial masses, metallicities and rotation
rates, as well as the convolution of yields from individual stars with an
observed distribution of rotational velocities. As far as binarity is concerned,
preliminary computations suggest that it might increase the \chem{26}{Al} yields
computed for single stars for WR initial
masses lower than about 40 M$_\odot$, the reverse effect being obtained for
more
massive stars. Clearly, this important question deserves further
studies. This is even more the case following recent observations of the
$\gamma^2$
Vel binary system.
 
\subsubsection{The $\gamma^2$ Vel system}

The presently available observations cannot help disentangling the relative
contributions of non-exploding WR stars, exploding ones (SNIb/Ic) and SNII to
the present \chem{26}{Al} galactic content. In this respect, a recent
observation of
high value provides an upper limit of about 10$^{-5}$ ph cm$^{-2}$s$^{-1}$ for the
1.8 MeV luminosity of the $\gamma^2$ Vel binary system containing an
O-type and a WR star (of WC8 subtype, Diehl et al. 1999). In order to evaluate the
compatibility of this observation with predictions, at least the initial mass
and  
metallicity of the WR progenitor, as well as the age of $\gamma^2$ Vel have to
be
known. The proximity of 
$\gamma^2$ Vel justifies the use of solar metallicity stellar models.
Non-rotating
single star models (Meynet et al. 1994), combined with the position of the O-star
component
in the HR diagram, lead to an age  between 3.5 and 4.5 My, and an initial WR
mass
between 40 and 60 M$_\odot$ (Schaerer et al. 1997). A 60 M$_\odot$ WR progenitor is
predicted to have a 1.8 MeV luminosity exceeding the observed upper limit by a
factor
of about 2, while the calculated luminosity lies below this limit in the case
of a
40 M$_\odot$ model star. From these considerations alone, one
can conclude that the observed upper limit is compatible with  
the predictions for $M \lsimeq 40$ M$_\odot$  single, non-rotating and
non-exploding
WR stars. Of course, $\gamma^2$ Vel is a binary system. As discussed above, a
$M
\gsimeq 40$ M$_\odot$ WR component cannot be excluded by the observation of
Diehl et al. (1999) if indeed a large enough fraction of its \chem{26}{Al}-loaded
ejecta
is accreted by its O-star companion.

\section {Summary and perspectives}

The decay of a variety of nuclides  with half-lives ranging from some years to
billions of years probably originating from outside the solar system are now recorded
in live or fossil form in the solar system. These findings bring a rich variety of
information about a vast diversity of highly interesting astrophysical questions,
some of which are briefly reviewed here with special emphasis on the contribution of
Jerry Wasserburg to this broadly interdisciplinary field of research.

First, the virtues of the
long-lived (half-lives $t_{1/2}$ close to, or in excess of
$10^9$ y) radionuclides as galactic chronometers are discussed in the light of recent
observational and theoretical works. It is concluded that the trans-actinide clocks
based on the solar system abundances of \chem{232}{Th}, \chem{235}{U} and
\chem{238}{U} or on the \chem{232}{Th} surface content of some old stars are still
unable to meaningfully complement galactic age estimates derived from
other independent astrophysical methods. In this respect, there is reasonable hope
that the \chem{187}{Re}-\chem{187}{Os} chronometric pair could offer better
prospects. The special case of \chem{176}{Lu}, which is a pure s-process product, is
also reviewed. It is generally considered today that this radionuclide cannot be
viewed as a reliable s-process chronometer.

Second, we comment on the astrophysical messages that could be brought by short-lived
radionuclides ($10^5 \lsimeq t_{1/2} \lsimeq 10^8$ y) that have been
present in live or in fossil form in the early solar system. From an astrophysical
point of view, the demonstrated early existence of live short-lived radionuclides is
generally considered to provide the most sensitive radiometric probe concerning
discrete nucleosynthetic events that presumably contaminated the solar system at
times between about $10^5$ and $10^8$ y prior to the isolation of the solar material
from the general galactic material. Of course, this assumes implicitly that the
radionuclides of interest have not been synthesized in the solar system itself. This
is still a matter of debate, as we briefly stress. If indeed the short-lived
radionuclides that have been present live in the early solar system are not of local
origin, the external contaminating agents that have been envisioned are supernovae,
evolved stars of the Asymptotic Giant Branch (AGB) type, or massive mass-losing
stars of the Wolf-Rayet (WR) type. We comment especially on some aspects of the AGB or
WR contamination. In the latter case, we discuss more specifically the role of
rotation and of binarity on the predicted yields of
\chem{26}{Al}, a radionuclide of special cosmochemistry and astrophysics interest.
Some comments are also devoted to
\chem{146}{Sm} and \chem{205}{Pb}. The former one is a short-lived p-process
radionuclide that has most probably been in live form in the solar system, while the
latter one is of s-process origin. It is shown to raise interesting nuclear physics
and astrophysics questions, and to deserve further cosmochemical studies in order to
evaluate its probability of existence in live form in the early solar system.

Third, the case of extinct short-lived radioactivities
carried by pre-solar grains is shortly mentioned, and some comments are
made about the possible origin of these grains. 

Finally, a brief mention is made to $\gamma$-ray line astrophysics, which provides
interesting information on live short-lived radionuclides in the present interstellar
medium, and thus complements in a very important way the study of extinct
radionuclides in meteorites. This is illustrated with the case of \chem{26}{Al}.
 
Jerry Wasserburg has written important
pages of many chapters of this `astrophysics of cosmic radioactivities' with
his incomparable dynamism and competence. He has without doubt
triggered much work in the field and shown many new ways.  For sure, it must be
most gratifying for him to know that so much remains to
be done and said, and that  each day brings its share of renewed excitement and
laboratory discoveries.
\vskip5truemm
\noindent{\it Acknowledgments} The authors are grateful to Larry Nittler and to an
anonymous referee for their very careful reading of the manuscript, and for their many
comments.  
\vfill\eject

\vfill\eject
\centerline{FIGURE CAPTIONS}
\vskip1.0truecm

\noindent {\bf Figure 1} Ages of the Universe ($T_{\rm U}$), of galactic
globular
 clusters 
($T_{\rm GC}$) and of the galactic disc ($T_{\rm disc}$).  Estimates from
cosmological models are specified for the standard hot Big Bang model by values of
$\Omega_0$, the ratio of the present density of the Universe to the critical density
that would make it `flat' (zero space curvature). Predictions are also displayed for
a non-standard flat model with a non-zero cosmological constant in terms of the
so-called deceleration parameter  $q_0$, which equals $3\Omega_0/2 - 1$ in the
considered model (e.g. Tayler 1986 for more details). In both cases, the
displayed ranges correspond to values $100 \geq H_0$(km/s/Mpc)$ \geq 50$ of the
Hubble constant $H_0$ at the present time. Adopted values of the redshift $z$ due to
the Hubble expansion are also given; these values are in no way meant to be precise.
The label SNIa defines the age limits
derived from values of those cosmological model parameters that are determined with
the use of SNIa as standardized candles. The  ages derived from HRD analyses of
globular clusters  or open clusters are in the ranges labelled GC and OC,
respectively. The predictions
based on white dwarf luminosity functions are noted WD, while the
nucleo-cosmochronological evaluations of  the age $T_{\rm nuc}$ of the nuclides
in the solar system are marked with  the label SS (from Arnould \& Takahashi 1999,
where more details and references can be found)
\vskip8truemm

\noindent {\bf Figure 2} Mass of \chem{26}{Al}, in solar units and as a
function of time (i.e. of the
thermal pulse number), predicted in the intershell layers of a \mass{2.5} solar
metallicity AGB star.
$^{26}$M$_{\rm H}$ (filled circles) is the total
\chem{26}{Al} mass  left over by the H-burning shell in those layers not
engulfed by
the next pulse convective tongue. That mass is directly brought to the surface
during
the dredge-up episodes. $^{26}$M$_{\rm He}$ (open circles) is
the total mass of
\chem{26}{Al} present in the C-rich layers left over by the pulse convective
tongue.
Part of this \chem{26}{Al} (up to 40\% according to
recent calculations by Mowlavi 1999; see also Goriely 1999) is eventually mixed into the envelope
during
the dredge-up episode. The production of primary
\chem{13}{C}, and the possible resulting destruction of a fraction of
$^{26}$M$_{\rm
He}$, is not included in the calculations. Each symbol corresponds to a thermal
pulse
\vskip8truemm

\noindent {\bf Figure 3} Masses of \chem{26}{Al} ejected by WR stars of
different initial masses and
metallicities $Z$ at the end of its WC-WO phase (from Meynet et al. 1997, referred
to as `present work') . The predictions of
$Z = 0.02$ WR models computed by Langer et al. (1995) are also displayed for
comparison
\vskip8truemm

\noindent {\bf Figure 4} Evolution of the total mass $M_{\rm TOT}$ and of the
mass of the convective core
$M_{\rm CONV}$ as a function of time for a non-rotating 60 $M_\odot$ model star
($\Omega = 0$), and for a star that just differs by its angular velocity
$\Omega =
0.6 \Omega_{\rm c}$, where $\Omega_{\rm c}$ is the break-up angular velocity.
The ordinate displays the mass $M$ at a given point inside the star normalized to the
mass of the Sun. In this scale $M = 0$ and $M = M_{\rm TOT}$ correspond to the stellar
center and surface, respectively.
Various evolutionary stages are indicated on the right at the corresponding values of
 $M_{\rm TOT}$
\vskip8truemm

\noindent {\bf Figure 5} Abundance ratios $\mbox{(R/S)}_0$ of various
radionuclides R relative to stable neighbors S for a 60 \ms$\,$ model star with $Z =
0.02$ versus
$\Delta^\ast$, interpreted as the period of free decay elapsed between the production
of the radionuclides and their incorporation into forming solar system solids. All
the displayed ratios are normalized to
$(\chim{107}{Pd}/\chim{108}{Pd})_0 = 2\times10^{-5}$ (e.g. Wasserburg 1985)
through the application of a common factor $d(\Delta^\ast)$, which may be seen as a
`dilution' factor measuring the fraction of the produced radionuclides that has been
effectively able to contaminate the solar system.
The values of this
factor are indicated on the Pd horizontal line for 3 values of
$\Delta^\ast$. Other available experimental data (labelled Exp) are displayed.
They are adopted from MacPherson et al. (1995) for Al, Srinivasan et al. (1994)
for
Ca (see also Sahijpal et al. 1998), Murty  et al. (1997) for Cl, and
Huey \& Kohman (1972) for Pb [see also
Chen \& Wasserburg (1987), who propose the somewhat larger value
(\chem{205}{Pb}/\chem{204}{Pb})$_0 \approx 3 \times 10^{-4}$ (not shown)] (see
\protect Arnould et al. (1997b) for more details)
\vskip8truemm

\noindent {\bf Figure 6}  Galactocentric radius dependence of the mass of
$^{26}$Al ejected by non-rotating non-exploding WR stars. The Galaxy is divided
into 15 concentric rings with 1 kpc width. More
computational details are provided in the main text  (taken from
Arnould et al. 1997)

\vfill\eject
%
%FIGURE 1
%
\begin{figure}[tb]
\centerline
{\bf
%{\vbox{\psfig{figure=coswasf0.ps,height=11.5cm,width=15.0cm}}}
{\vbox{\psfig{figure=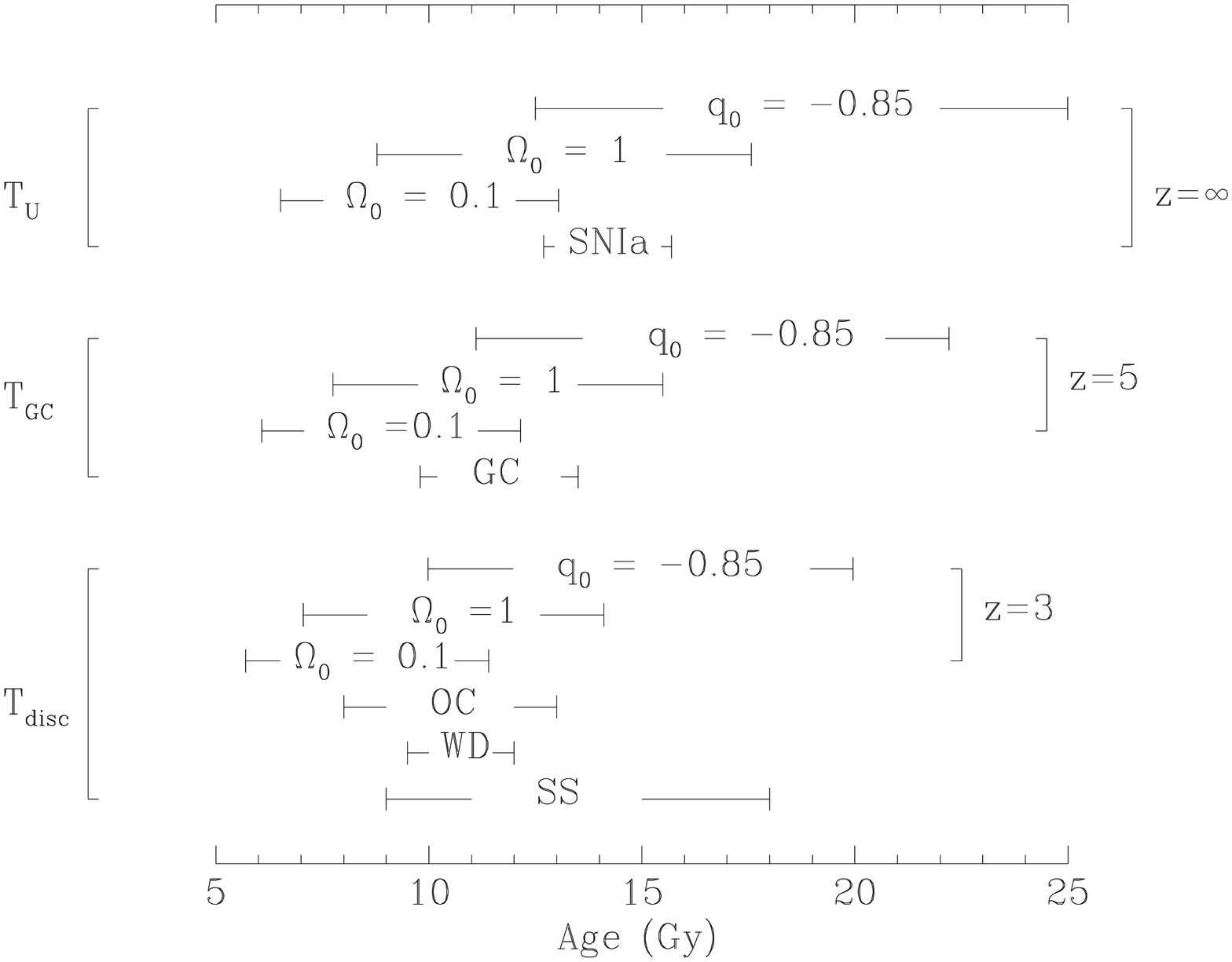,height=12.3cm,width=16.0cm}}}
}
\caption[ ]{\small }
\end{figure}
\newpage
%
% FIGURE 2
%
\begin{figure}[tb]
\centerline
{\bf
%{\vbox{\psfig{figure=agbAl26.eps,height=8.2cm,width=15.0cm}}}
{\vbox{\psfig{figure=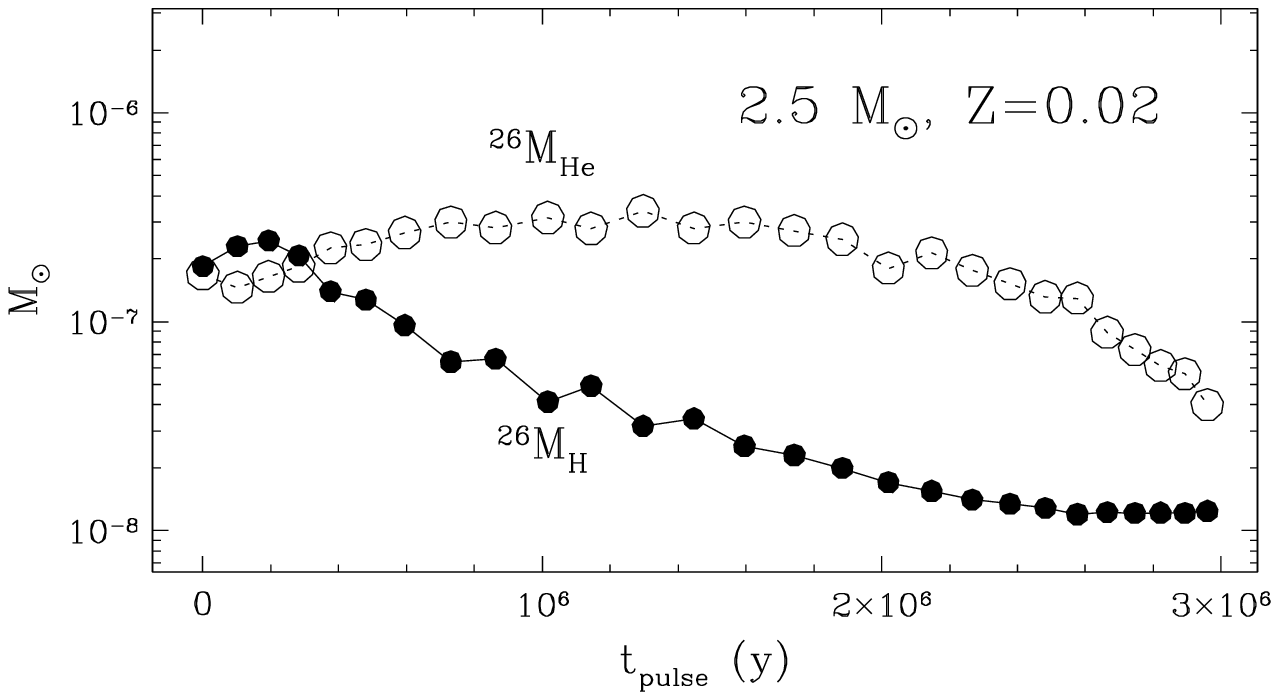,height=8.75cm,width=16.0cm}}}
}
\caption[ ]{\small }
\end{figure}
\newpage
%
% FIGURE 3
% 
\begin{figure}[tb]
\centerline
{\bf
{\vbox{\psfig{figure=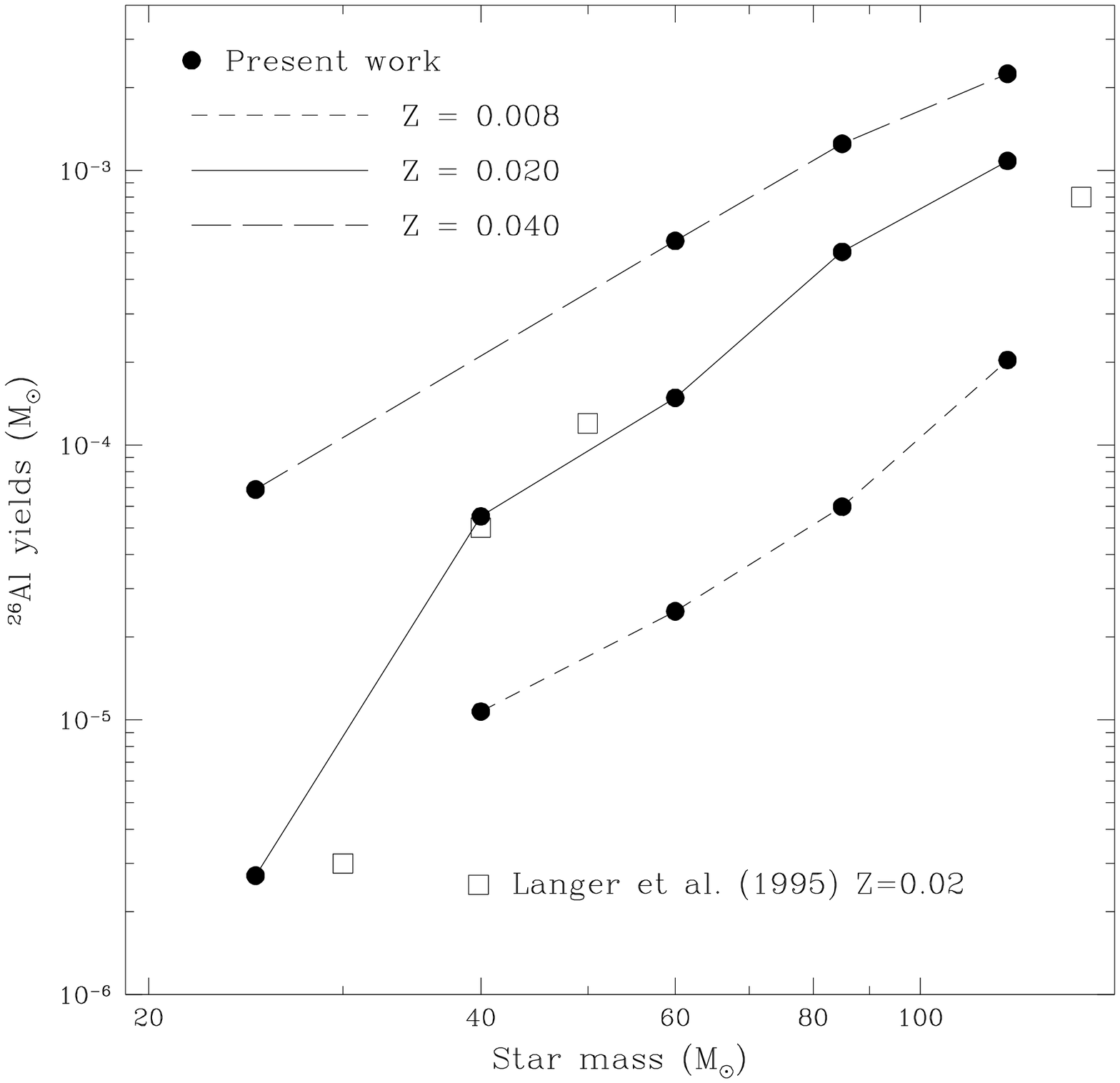,height=13.8cm,width=15.0cm}}}
}
\caption[ ]{\small }
\end{figure}
\newpage
%
% FIGURE 4
%
\begin{figure}[tb]
\centerline
{\bf
%{\vbox{\psfig{figure=coswasf3.ps,height=16.0cm,width=\textwidth}}}
{\vbox{\psfig{figure=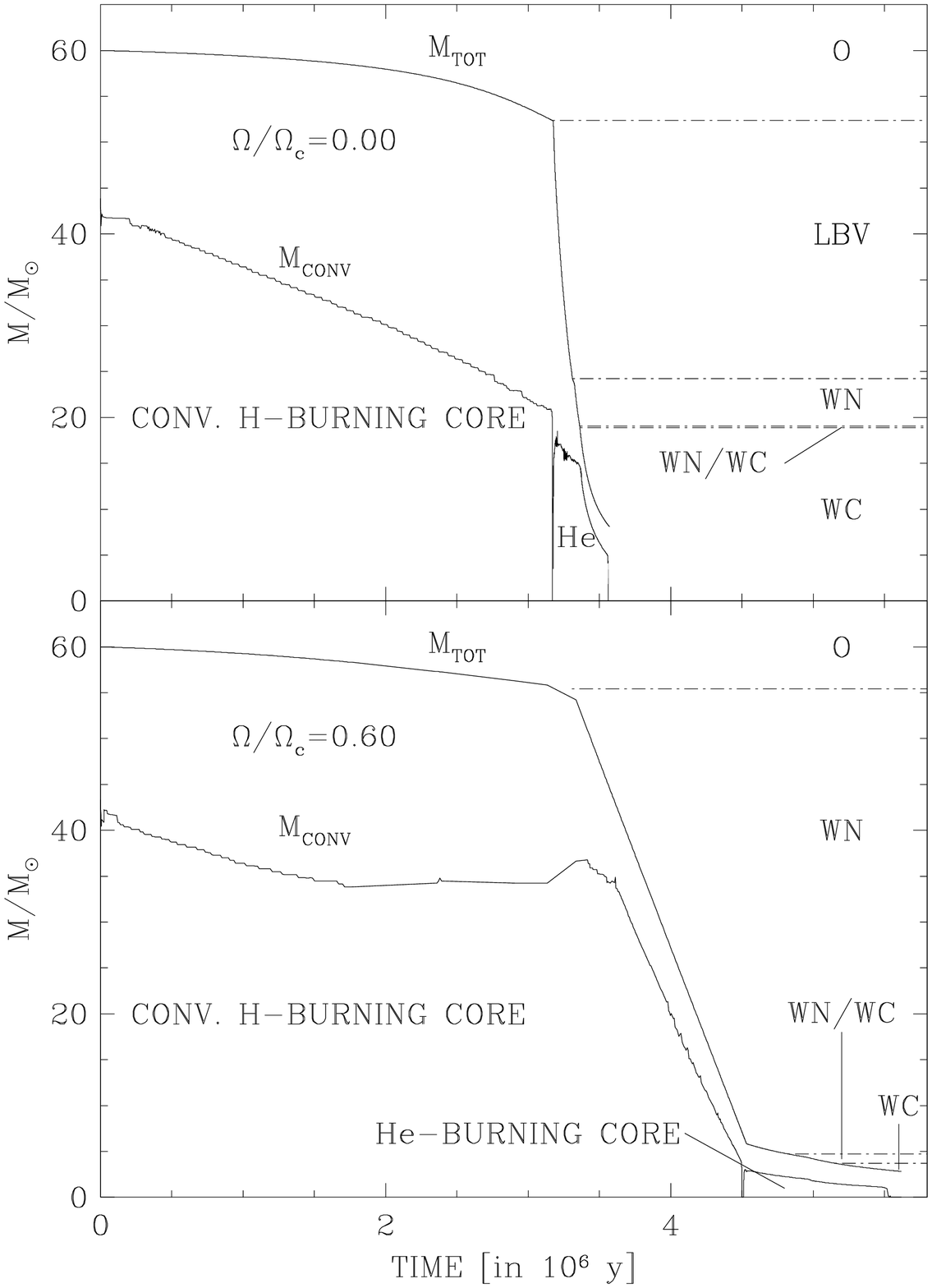,height=20.7cm,width=15.0cm}}}
}
\caption[ ]{\small }
\end{figure}
\newpage
%
% FIGURE 5
%
\begin{figure}[tb]
\centerline
{\bf
{\vbox{\psfig{figure=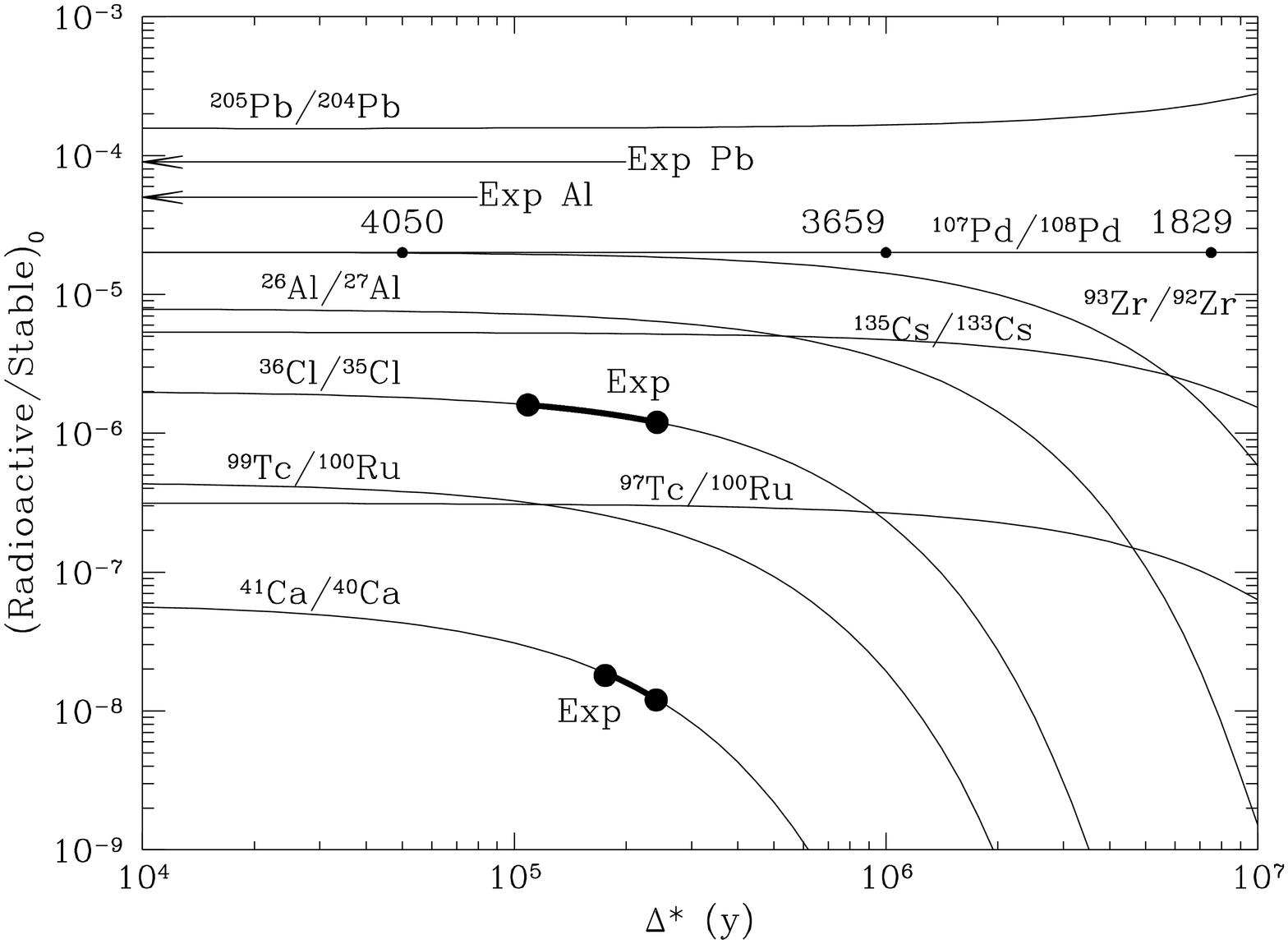,angle=270,height=10.4cm,width=15.0cm}}}
}
%\special{psfile=cosmicf1.ps angle=270 hscale=52 vscale=54 voffset=300}
\caption[]{\small }
\end{figure}
\newpage
%
% FIGURE 6
%
\begin{figure}[tb]
\centerline
{\bf
{\vbox{\psfig{figure=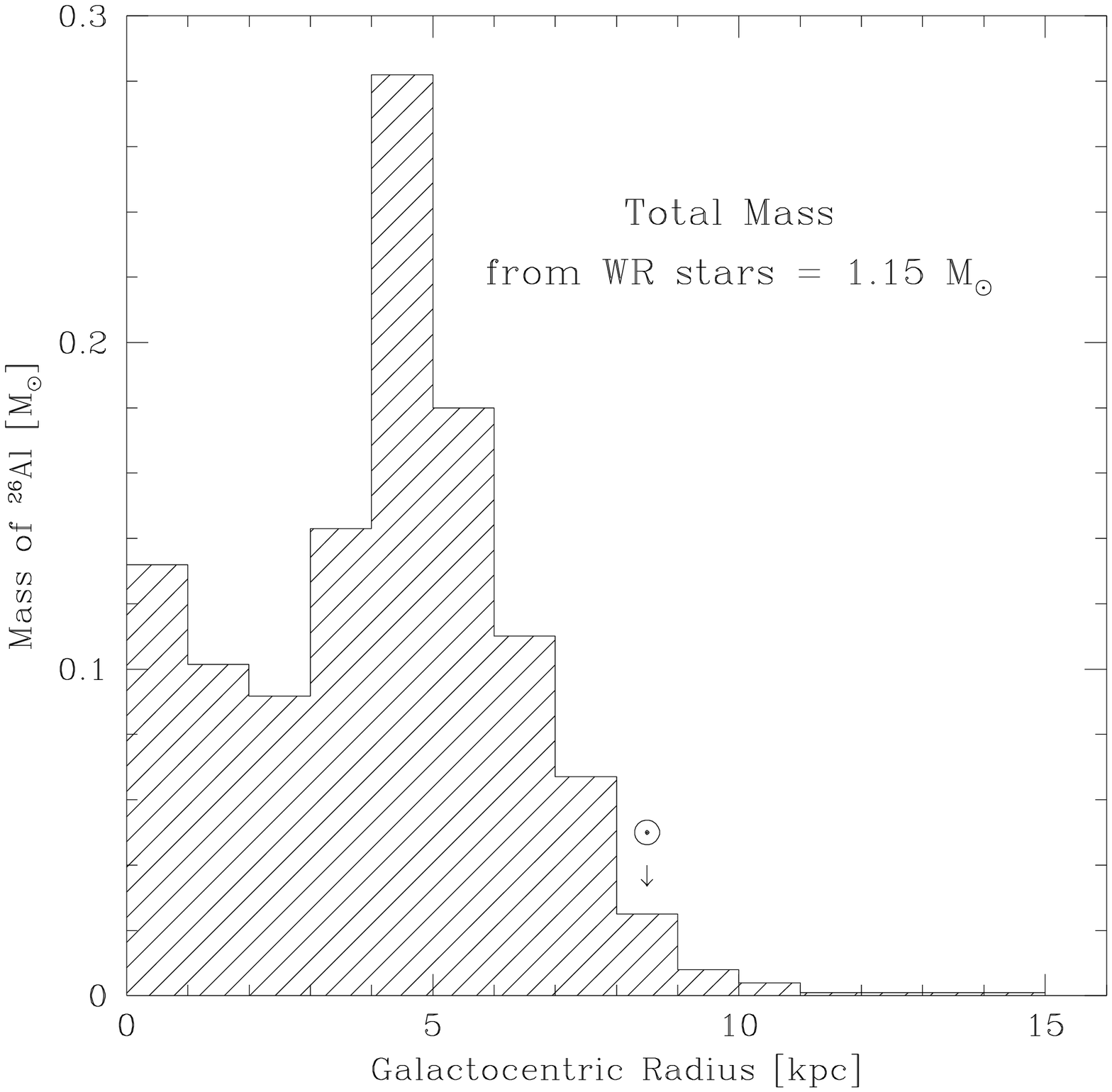,height=12.5cm,width=15.0cm}}}
}
\caption[ ]{\small }
\end{figure}
 
\end{document}